\documentclass[sigconf]{acmart}

\copyrightyear{2025}
\acmYear{2025}
\setcopyright{acmlicensed}
\acmConference[ASIA CCS '25]{ACM Asia Conference on Computer and Communications Security}{August 25--29, 2025}{Hanoi, Vietnam}
\acmBooktitle{ACM Asia Conference on Computer and Communications Security (ASIA CCS '25), August 25--29, 2025, Hanoi, Vietnam}\acmDOI{10.1145/3708821.3733881}
\acmISBN{979-8-4007-1410-8/2025/08}

\usepackage{algorithmic}
\usepackage{graphicx}
\usepackage{textcomp}
\usepackage{xcolor}
\usepackage{xspace}
\usepackage{algorithm}
\usepackage{subfigure}
\usepackage{multirow}
\usepackage{url}
\usepackage{bbding}
\usepackage{pifont} 
\usepackage{array}
\usepackage{diagbox}
\usepackage{float}
\usepackage{caption}
\usepackage{booktabs,caption}
\newcommand{\rev}[1]{{\color{black} #1}}
\newcommand{\revv}[1]{{\color{black} #1}}

\newcommand{\ours} {\textsc{ClearMask}\xspace}
\newcommand{\ourss}{\textsc{LiveMask}\xspace}

\begin{document}

\title{\ours: Noise-Free and Naturalness-Preserving Protection Against Voice Deepfake Attacks}
\author{Yuanda Wang}  
                                  % ...
\email{wangy208@msu.edu} 
\affiliation{
  \institution{Michigan State University}
  \city{East Lansing}
  \state{Michigan}
  \country{USA}  
  }

\author{Bocheng Chen}  
                                  % ...
\email{chenboc1@msu.edu} 
\affiliation{
  \institution{Michigan State University}
  \city{East Lansing}
\state{Michigan}
\country{USA}  
  }
\author{Hanqing Guo} 

\email{guohanqi@hawaii.edu}
\affiliation{
  \institution{University of Hawaii at Mānoa}
  \city{Honolulu}
  \state{Hawaii}
  \country{USA}
  }
\author{Guangjing Wang}  
                                  % ...
\email{guangjingwang@usf.edu} 
\affiliation{
  \institution{University of South Florida}
  \city{Tampa}
  \state{Florida}
  \country{USA}  
  }
\author{Weikang Ding}  
                                  % ...
\email{dingweik@msu.edu} 
\affiliation{
  \institution{Michigan State University}
  \city{East Lansing}
  \state{Michigan}
\country{USA}  
  }
\author{Qiben Yan}  
                                  % ...
\email{qyan@msu.edu} 
\affiliation{
  \institution{Michigan State University}
  \city{East Lansing}
  \state{Michigan}
  \country{USA}  
  }
  
\renewcommand{\shortauthors}{Yuanda Wang, Bocheng Chen, Hanqing Guo, Guangjing Wang, Weikang Ding \& Qiben Yan}

\begin{abstract}
Voice deepfake attacks, which artificially impersonate human speech for malicious purposes, have emerged as a severe threat. 
Existing defenses typically inject noise into human speech to compromise voice encoders in speech synthesis models.
However, these methods degrade audio quality and require prior knowledge of the attack approaches, limiting their effectiveness is diverse scenarios.
Moreover, real-time audios, such as speech in virtual meetings and voice messages, are still exposed to voice deepfake threats.
To overcome these limitations, we propose \ours, a noise-free defense mechanism against voice deepfake attacks. 
Unlike traditional approaches, \ours modifies the audio mel-spectrogram by selectively filtering certain frequencies, inducing a transferable voice feature loss without injecting noise.
We then apply audio style transfer to further deceive voice decoders while preserving perceived sound quality. 
Finally, optimized reverberation is introduced to disrupt the output of voice generation models without affecting the naturalness of the speech. 
Additionally, we develop \ours, named \ourss, to protect streaming speech in real-time through universal frequency filter and reverberation generator.
Our experimental results show that \ours and \ourss effectively prevent voice deepfake attacks from deceiving speaker verification models and human listeners, even for unseen voice synthesis models and black-box API services.
Furthermore, \ours demonstrates resilience against adaptive attackers who attempt to recover the original audio signal from the protected speech samples.
\end{abstract}


\maketitle
\section{Introduction}
With the rapid advancement of deep learning, voice synthesis models have become increasingly powerful, producing speech sounds highly natural and lifelike~\cite{elevenlabs,playht}.
Building on it, state-of-the-art voice synthesis models can create convincing speech content using a short speech sample as a reference, replicating not only the speaker's voice but also their prosody and rhythm to make the output indistinguishable from real human speech~\cite{wang2018style}.
However, these advancements also lead to threats when generated speech is misused, known as voice deepfake attacks~\cite{wenger2021hello}.
In 2023, fraudsters successfully bypassed bank authentication systems using synthetic voices~\cite{breakbank}, and even orchestrated a fake kidnapping by synthesizing a girl's voice~\cite{kidnapping}. 
In another instance, criminals deceived a company into transferring \$200,000 using fake speech~\cite{scammoney}. 
Some individuals have exploited this technology to generate hate speech using the voices of celebrities, causing significant reputational harm~\cite{voiceabuse}. 
Additionally, synthetic voices can be used to compromise voice assistants during user verification processes or to execute malicious voice commands~\cite{wang2023practical}. 
These examples underscore the risks of synthetic speech being used to deceive humans or automatic speaker verification (ASV) systems. 
Unfortunately, preventing the misuse of synthetic speech by restricting speech synthesis models remains a substantial challenge. 
As a result, exposing unprotected voices on public media platforms raises serious threats to user security and privacy.

\begin{table}[tbp]
\centering
\captionsetup{aboveskip=2pt, belowskip=-5pt}
\small
\caption{Comparison of \ours with existing defenses.}
\label{tab:compare}
\begin{tabular}{|c|c|c|c|c|} 
\hline
\begin{tabular}[c]{@{}c@{}}Defenses\end{tabular}   & \begin{tabular}[c]{@{}c@{}}Effective-\\ness\end{tabular} & \begin{tabular}[c]{@{}c@{}}Trans-\\ferability\end{tabular} & \begin{tabular}[c]{@{}c@{}}High-\\quality\end{tabular} & \begin{tabular}[c]{@{}c@{}}Real-\\time\end{tabular}  \\ 
\hline
Attack-VC~\cite{huang2021defending}  &Medium&Low&Low&Low      \\ 
\hline
SampleMask~\cite{liu2023protecting} &Low&Medium&Medium&High  \\ 
\hline
VSMask~\cite{wang2023vsmask}     &High&Low&Low&High       \\ 
\hline
AntiFake~\cite{yu2023antifake}   &High&High&Medium&Low     \\ 
\hline
\textbf{\ours}       &   \textbf{High}&    \textbf{High}&  \textbf{High}&  \textbf{High}        \\
\hline
\end{tabular}
\vspace{-15pt}
\end{table}

One common defense against artificially generated speech is liveness detection~\cite{ahmed2020void}, which identifies unnatural speech not originate from human vocal tracts. 
However, this approach is limited as it cannot prevent attacks designed to deceive human listeners. 
To address the limitation, Attack-VC~\cite{huang2021defending} mitigates voice deepfake attacks by masking speech audio before it is uploaded to public social media platforms.
By injecting carefully optimized perturbations into audio signals, Attack-VC prevents attackers from generating convincing speech samples capable of deceiving ASV systems or human listeners. 
\rev{Despite its effectiveness in specific scenarios, achieving a comprehensive defense in diverse real-world applications remains challenging. 
These challenges can be categorized into four key aspects. 
First, achieving a high success rate is critical for all defense mechanisms. Even a small chance of successful attacks can compromise the defense, leading to severe and unacceptable damages.
Second, defenders often lack prior knowledge about the specific voice synthesis models that attackers employ. 
Therefore, existing white-box defenses cannot be generalized to protect against various voice synthesis models.
Third, the protected speech audio must retain clarity and intelligibility to ensure its usability in practical applications, such as videos or voice messages.
Fourth, safeguarding real-time speech audio in instant communication applications, e.g., FaceTime~\cite{facetime} and online meetings, is essential. 
These scenarios, however, require protection mechanisms to function with minimal latency, which presents a technical challenge.}

\rev{
Consequently, we identify four essential features  for effective protection against voice deepfake attacks: \textit{effectiveness, transferability, high-quality}, and \textit{real-time}, corresponding to the four challenges outlined above. 
In Table~\ref{tab:compare}, we evaluate existing defenses across these dimensions. 
However, while individual defenses demonstrate unique strengths, none of them satisfy all four requirements.
Attack-VC~\cite{huang2021defending} and VSMask~\cite{wang2023vsmask} are white-box defenses with limited transferability. SampleMask~\cite{liu2023protecting} requires querying a specific model, and demonstrates limited effectiveness. 
While AntiFake~\cite{yu2023antifake} exhibits strong effectiveness and transferability, it suffers from poor audio quality and lacks real-time feasibility.}
\rev{
To overcome these challenges, we propose \ours, a noise-free defense against voice deepfake attacks. 
While ensuring \textit{effectiveness}, we adopt an ensemble encoder approach to enhance the \textit{transferability}. 
Moreover, unlike traditional methods which inject noise to mask voice samples, \ours leverages multiple natural sound effects to generate \textit{high-quality} speech. 
Specifically, it first modifies the input mel-spectrogram of voice synthesis models by filtering out specific frequencies in speech spectrograms. 
Next, \ours employs audio style transfer to obscure distinctive voice features, effectively misleading deepfake voice generation. 
Finally, it introduces a well-optimized room impulse response (RIR) to create unique reverberation effects that further enhance the protection.
In addition, we design \ourss, a \textit{real-time} mode of \ours designed for online speech protection.
\ourss employs universal frequency filter and reverberation sound effect to provide immediate protection for instant communication scenarios.}

In the evaluation, we test \ours and \ourss on both open-source and commercial voice synthesis platforms. 
The results demonstrate that \ours and \ourss effectively prevent unauthorized voice synthesis in both offline and online scenarios. 
With this protection, the generated deepfake voices fail to deceive ASV systems or human listeners.
Furthermore, \ours demonstrates robustness against adaptive attackers with varying capabilities who attempt to remove the reverberation, effectively maintaining its defensive performance.
Overall, our contributions are summarized as follows: 
\begin{itemize}
    \item  We introduce \ours, a noise-free protection mechanism against malicious voice cloning. \ours employs spectrogram masking to modify the mel-spectrogram. In addition, we utilize audio style transfer and artificial reverberation to further obscure distinctive voice features. 
    \item We propose \ourss, a real-time mode of \ours designed for immediate protection. It applies a general frequency filter and universal reverberation generation to protect streaming online speech with millisecond-level latency.
    \item We provide a comprehensive system design to optimize the three stages of \ours. Through a surrogate ensemble encoder, we strike a balance among effectiveness, transferability and audio quality.
    \item We comprehensively evaluate the protection performance of \ours and \ourss on five voice synthesis models including commercial APIs. Our results demonstrate that \ours achieves effective, transferable, high-quality, and real-time protection in various scenarios.
\end{itemize}

\vspace{-5pt}
\section{Preliminaries}
\subsection{Voice Deepfake Attacks and Defenses}
\begin{figure}[t]
    \centering
    \captionsetup{aboveskip=5pt, belowskip=-18pt} 
    \includegraphics[width=0.45\textwidth]{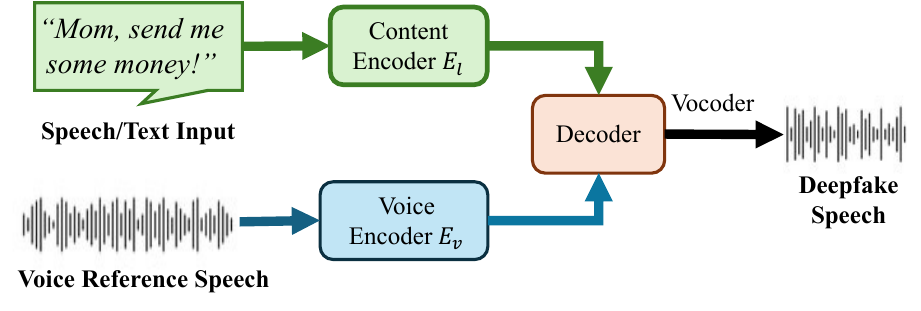}
    \caption{A general framework of voice synthesis models.}
    \label{fig:1_deepfake_generation}
\end{figure}

Voice synthesis models produce artificial speech by integrating linguistic content derived from source speech or textual input with voice features extracted from a reference speech sample.
Voice synthesis approaches are typically categorized into Voice Conversion (VC), which modifies the voice of a given speech sample to mimic a target speaker~\cite{chou2019one, chen2021again, qian2019autovc}, and Text-to-Speech (TTS), which generates speech from text using the target speaker's voice~\cite{casanova2022yourtts}.
Fig.~\ref{fig:1_deepfake_generation} illustrates a general framework of voice synthesis, where content and voice features are encoded into embeddings, decoded to produce a mel-spectrogram, and converted into audible waveforms. 
VC usually creates more expressive results by retaining human emotion and intonation, while advanced TTS models also achieve high realism with style embedding~\cite{wang2018style}.
To prevent unauthorized voice synthesis, users can employ adversarial examples to compromise the voice encoder module $E_v$, leading to unqualified synthetic voice samples.
The optimization process of this adversarial speech is:
\begin{equation}
arg\mathop{\textnormal{max}}\limits_{\delta} \ \Vert E_v(\boldsymbol{x}_{r}+\delta) - E_v(\boldsymbol{x}_r) \Vert_2 , \quad \textbf{s.t.} \ \Vert \delta \Vert_\infty < \epsilon,
\end{equation}
where $\boldsymbol{x}_{r}$ is the victim's speech sample and $\delta$ is perturbation signal.
However, this approach heavily depends on gradient back-propagation from a single encoder model, which limits its transferability in black-box scenarios.
Additionally, the perturbations $\delta$ degrade the audio quality of the adversarial speech samples. 
Therefore, more advanced methods are needed to effectively protect speech against various attack techniques while preserving audio quality and naturalness.

\vspace{-5pt}
\subsection{Audio Style Transfer}
Inspired by image style transfer~\cite{gatys2016image}, audio style transfer has emerged with the goal of modifying the texture of sound, such as the timbre of a musical instrument~\cite{grinstein2018audio, steinmetz2022style}.
Both audio style transfer and voice conversion focus on embedding features of one audio sample into another to synthesize a new audio.
An audio style transfer model generally includes a style extraction module $\boldsymbol{E}_s$ and a style synthesis module $\boldsymbol{S}$, and can be mathematically represented as follows:
\begin{equation}
    \boldsymbol{x}'_{t} = \boldsymbol{S}(\boldsymbol{x}_{t}, V_{s}), \quad \textbf{s.t.}~V_{s} = \boldsymbol{E}_s(\boldsymbol{y}_{t}),
\end{equation}
where $\boldsymbol{x}_{t}$ is the source audio and $\boldsymbol{y}_{t}$ is the style reference audio.
Although audio style transfer does not affect the speech content features, the difference in sound texture could mislead the voice encoder in voice synthesis models and damage the extracted voice embedding vectors.

\vspace{-10pt}

\subsection{Reverberation in Adversarial Examples}\label{Sec: Reverb}
Reverberation is a phenomenon caused by multi-path effects in the physical world.
The reverberation can be quantified using RIR, defined as the response signal after a pulse is played in a given environment.
We can simulate reverberation by convolving an RIR with an original audio $\boldsymbol{x}_{t}$:
\begin{equation}
    \boldsymbol{x}^{R}_{t} = \boldsymbol{x}_{t} \ast h,
\end{equation}
where $h$ is the reversed RIR signal, and $\boldsymbol{x}^{R}_{t}$ is the audio with reverberation effect.

While reverberation in common environments is often too subtle to affect human perception, it plays a critical role in adversarial audio attacks in the physical world.
For instance, when attackers replay adversarial audio examples to compromise speech recognition models, the typically weak perturbations can result in attack failure due to reverberation distortion.
To address this challenge, attackers must measure the RIR at the precise locations of the adversarial audio transmitter and victim microphone, then apply the RIR filter to the original speech to ensure the target device receives the intended perturbation~\cite{schonherr2020imperio, chen2020metamorph}.
\vspace{-3pt}
\section{Threat Model}\label{Sec: Threatmodel}
\begin{figure}[t]
    \centering
    \captionsetup{aboveskip=3pt, belowskip=-0pt} 
    \subfigure[Voice deepfake attacks scenarios.]{\includegraphics[width=0.45\textwidth]{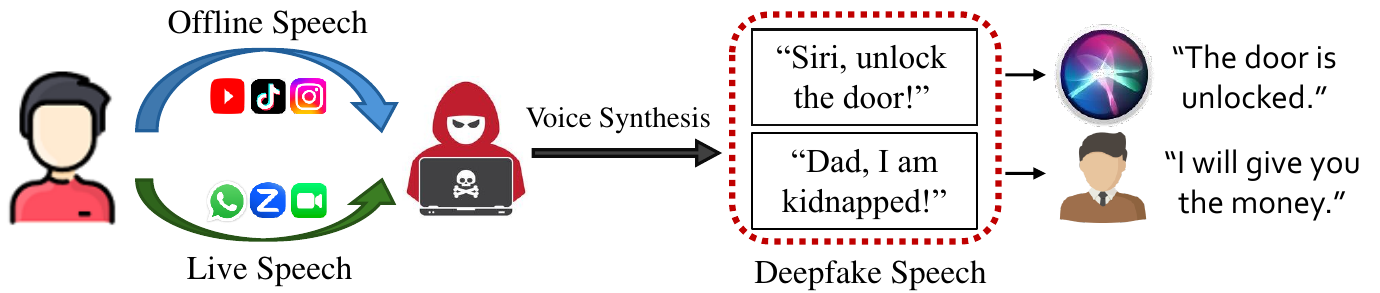}
    \label{fig: unprotected}}    
    \subfigure[With \ours and \ourss protection, the synthesized speech could no longer deceive ASV or human ears.]{\includegraphics[width=0.45\textwidth]{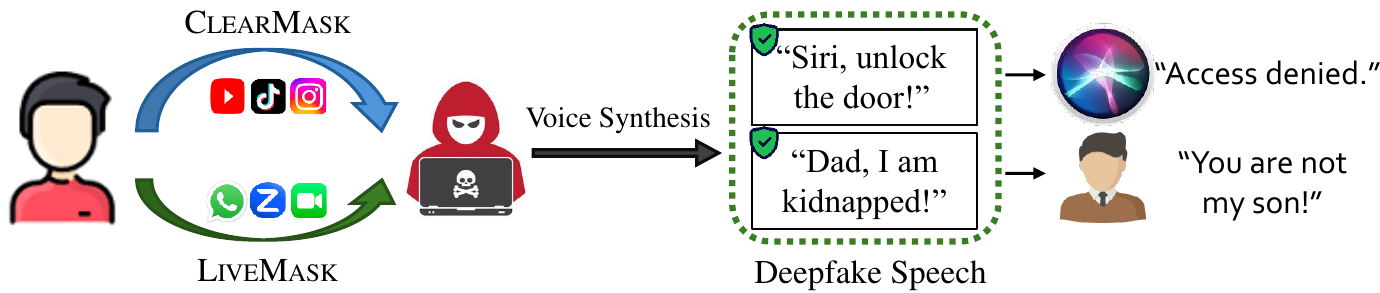}
    \label{fig: protected}}    

    \caption{Compared with raw speech, \ours protected speech is robust against voice deepfake attacks.}
        \vspace{-15pt}
    \label{fig: application scenario}
\end{figure}

\begin{figure*}[tbp]
    \centering
    \includegraphics[width=0.80\textwidth]{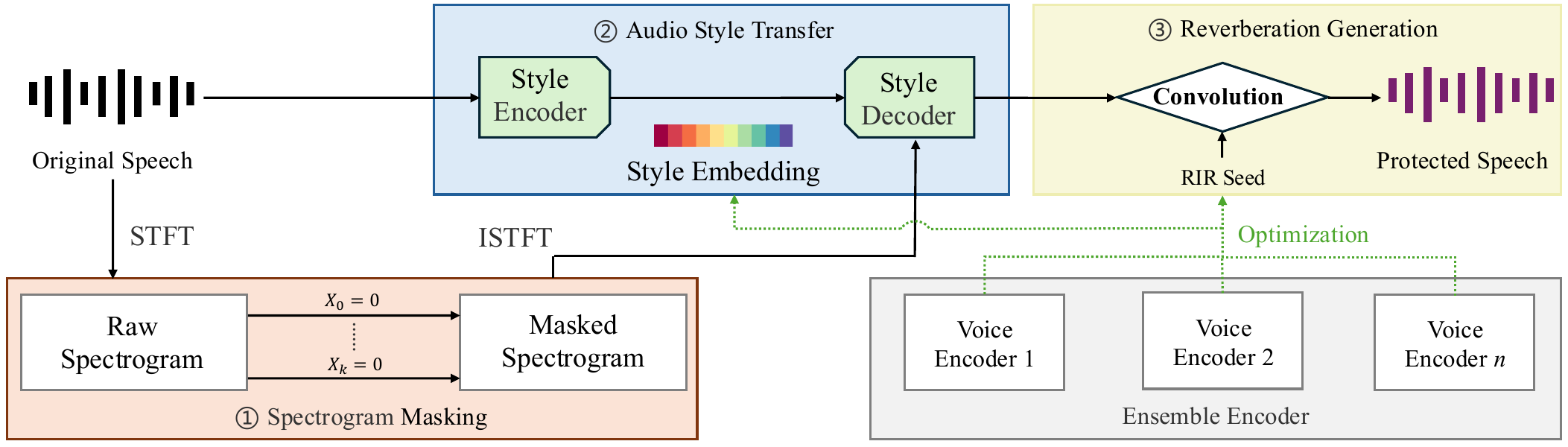}
    \caption{System overview of \ours. It first applies spectrogram masking to modify the mel-spectrogram. Next, an ensemble voice encoder is leveraged to optimize audio style transfer and reverberation generation stages, improving the defense effectiveness and transferability.}
    \label{fig:2_system_design}
    \vspace{-10pt}
\end{figure*}
Current state-of-the-art voice cloning models are capable of synthesizing high-quality speech using just a few seconds speech sample as a reference, making it particularly challenging to effectively defend against such attacks. 
In this section, we define the threat model of voice deepfake attacks, elaborating on the attacker’s knowledge and methods for acquiring reference voice samples, as well as the defender’s capabilities of mitigating potential voice deepfake threats.

\vspace{-8pt}
\subsection{Adversary Capability} 
\noindent{\textbf{Approaches:}}
As illustrated in Fig.~\ref{fig: unprotected}, people commonly expose their voices in two scenarios.
On one hand, \textit{offline speech}, such as videos shared on public social media platforms like Instagram, TikTok, and YouTube, can be exploited by attackers who download these videos and use the speech to impersonate the user’s identity.
On the other hand, \textit{online speech} from real-time applications, such as online meetings and phone calls, also puts the user’s voice at risk. 
Protecting speech in online scenarios is particularly challenging, as users lack sufficient time to optimize the masking of these speech samples.
Once an attacker successfully clones a victim’s voice, the synthesized speech can be used to control voice assistants or deceive human listeners, leading to potentially severe consequences such as financial losses or physical threats~\cite{wang2022ghosttalk}.

\noindent{\textbf{Knowledge:}}
Adversaries are capable of synthesizing deepfake speech by leveraging readily available open-source speech synthesis models or commercial APIs.
Reference speech can be obtained from public social platforms or recorded from live-streaming audio.
Moreover, attackers may possess private information about victims, such as their names and contact details, enabling them to execute more sophisticated attacks, including spam phone calls.
Additionally, if attackers become aware that speech samples have been protected, they may attempt to bypass the protective measures and recover the raw speech.
Alternatively, they could synthesize the victim's voice using various models and select the one with highest performance.

\vspace{-8pt}

\subsection{Defender Capability}
\noindent{\textbf{Approaches:}}
Fig.~\ref{fig: protected} illustrates the application scenarios of \ours.
For offline speech, defenders can utilize \ours, which leverages noise-free sound effects to safeguard speech prior to uploading it onto public social media platforms.
In online scenarios, defenders can activate \ourss, the real-time mode of \ours to enable fast protection.
With \ours protection, neither ASV models nor human listeners can identify the synthetic voice as belonging to the victim, thereby rendering voice deepfake attacks ineffective.

\noindent{\textbf{Knowledge:}}
In this work, we consider \ours a \rev{\textit{black-box}} defense, meaning that defenders have no prior knowledge about the adversaries.
Specifically, the model being used, including its architecture, weights, and training data, remains unknown.
Furthermore, they are prohibited from querying black-box models.
During the defense stage, defenders employing \ours can utilize public speech and RIR datasets to enhance protection performance.
Moreover, they have unrestricted access to open-source voice synthesis models and their well-trained checkpoints, allowing them to build a surrogate model for the transferable defense.

\vspace{-7pt}
\section{Methodology}
To address the technical challenges in existing defense approaches, including transferability, audio quality, and real-time feasibility, we design \ours, a novel speech protection method to prevent malicious voice synthesis.
The framework of \ours is illustrated in Fig.~\ref{fig:2_system_design}.
Instead of adding noise to the original speech, \ours implements three techniques to mask the real voice features: spectrogram masking, audio style transfer, and reverberation generation.
These noise-free methods allow us to reduce audio distortion and maintain naturalness for human perception.
\vspace{-10pt}
\subsection{Spectrogram Masking}
The high sampling rate of digital audio complicates direct time-domain audio signal processing. 
As a result, speech synthesis models utilize spectrograms as the input. 
Typically, audio waveforms are converted into spectrograms using Short-Time Fourier Transform (STFT), which are then mapped to the mel scale—a perceptual scale that aligns with human auditory sensitivity by emphasizing lower frequencies. 
During speech generation, the decoder outputs a mel spectrogram, which is converted into an audible waveform by vocoders such as HiFiGAN~\cite{kong2020hifi}.

Therefore, we propose masking the input mel-spectrogram of voice synthesis models to influence the output of the voice encoder $E_v$.
Existing work~\cite{liu2023protecting} has demonstrated that masking certain frequencies in the spectrogram can mislead the voice feature extraction process in speech synthesis, resulting in degraded synthetic speech while preserving overall audio quality.
However, this straightforward approach is not consistently effective across diverse speech samples.
Unlike methods that mask wide frequency bands, \ours sets only a few selected frequencies to zero to minimize audio quality degradation.
Furthermore, to reduce computational complexity, we first identify the frequencies with substantial power ($\geq \tau_p$), as audio spectrograms are inherently sparse matrices.
Next, we apply a greedy algorithm to select the frequencies to be masked, as detailed in Algorithm~\ref{Algorithm:frequency}.
\begin{algorithm}[htbp]
    \caption{Greedy Frequency Selection}\label{Algorithm:frequency}
    \begin{algorithmic}[1]
        \STATE \textbf{Input:} Spectrogram $\mathbf{X} = [X_0, X_1, \ldots, X_n]^{T}$
        \FOR {$X_i \in \mathbf{X}$}
            \IF {$\| X_i \|_{2} \geq \tau_p$}
                \STATE $\mathbf{X}_{\text{temp}} \gets \mathbf{X}$ \textbf{with} $X_i$ \textbf{set to} $0$
                \STATE $\Delta_i \gets \left\| \text{Mel}(\mathbf{X}) - \text{Mel}(\mathbf{X}_{\text{temp}}) \right\|_{2}$
            \ENDIF
        \ENDFOR
        \STATE $\mathcal{S} \gets$ indices of the top $k$ values of $\Delta_i$
        \STATE $\mathbf{X}' \gets \mathbf{X}$ \textbf{with} $X_{i \in \mathcal{S}}$ \textbf{set to} $0$
        \RETURN $\mathbf{X}'$
    \end{algorithmic}
\end{algorithm}


The greedy frequency selection method first ranks the frequencies based on the mel-spectrogram loss $ \Vert \text{Mel}(\mathbf{X}) - \text{Mel}(\mathbf{X}_{temp}) \Vert_{2}$, where $\mathbf{X}_{temp}$ represents the spectrogram after removing a specific frequency component $X_i$.
Next, we filter out only $k$ frequencies with the strongest impact on the input mel-spectrogram, striking a balance between audio quality and protection performance.
Notably, since the mel-spectrogram deviation serves as the loss function and does not depend on gradient back-propagation from any specific voice encoder model, this approach achieves transferable performance across diverse models.

\vspace{-10pt}

\subsection{Audio Style Transfer}\label{Section: style transfer}
While masking certain frequencies can protect voice features without noticeably degrading audio quality, voice patterns can still be extracted.
Therefore, additional methods are necessary to improve the effectiveness of protection.
Audio style transfer, which modifies sound texture without introducing noise, is another key technique that can be utilized to spoof voice synthesis models.
However, optimizing the audio style of input speech samples to achieve effective defense is challenging.
\rev{Directly using another speaker's speech as $\boldsymbol{y}_t$ to process the original speech is not feasible, as explained in Appendix~\ref{sec:voice_and_style}.}
In this section, we will illustrate the audio style optimization strategy for \ours protection.

\vspace{-2pt}
\subsubsection{Ensemble Encoders}
Existing defense methods that employ adversarial examples to attack voice synthesis models usually rely on gradient back-propagation.
However, due to overfitting, the transferability of defenses is limited when the gradient from a single encoder is used to optimize the adversarial example.
If the defense assumes only white-box scenarios, its effectiveness cannot be guaranteed when attackers employ alternative models to synthesize speech.
Therefore, it is crucial to ensure that protected speech examples remain effective across models with varying architectures.

Fortunately, although different voice encoders are employed, various voice synthesis models share the common goal of extracting the unique vocal patterns of a single speaker from speech samples with diverse content.
Regardless of differences in model architectures, all voice encoder modules extract the same voice features, even though they are represented differently.
Based on this prior knowledge, it is possible to design a transferable protection method that remains effective across different voice synthesis models.
However, variations in training data and model architectures inevitably result in different gradient values.
To achieve optimal performance across diverse models, \ours incorporates encoders with varying architectures and optimizes the loss function as follows:
\begin{equation}
    \mathcal{L}_{spk}(x_{t}^{\mathcal{M}}, x_{t}) = \sum_{i=0}^{n} \lambda_{i}\Vert E_{v}^{i}(x_{t}^{\mathcal{M}}) - E_{v}^{i}(x_{t}) \Vert_{2}, 
\label{equation: speaker encoder}
\end{equation}
where $E_{v}^{i}$ is the $i$-th encoder, and $x_{t}^{\mathcal{M}}$ is the masked speech sample generated by \ours.
Additionally, as the dimensions of the voice embedding vectors are different, we use $\lambda_{i}$ to adjust the weight of each encoder.
This loss function based on the ensemble encoder can significantly enhance protection transferability since the optimization is guided by gradients from a diverse collection of voice encoders.

\subsubsection{Style Optimization}
By leveraging the surrogate ensemble encoder, we can optimize the target audio style to modify the input speech and compromise different voice synthesis models.
In \ours, we apply DeepAFx-ST~\cite{steinmetz2022style} as the audio style transfer tool.
However, the challenge in the optimization process is that audio style transfer functions serve as a black box for voice encoder models, meaning that we cannot directly derive gradients to optimize the style embeddings.
To address the challenge, we need to query the surrogate model and find the optimal target audio style for the input speech.

Moreover, in audio style transfer process, the style embedding vector $\mathbf{V}$ is normalized to have a fixed $l_2$ norm.
Therefore, we can only flip each dimension without changing the value.
We set the input of audio style transfer to be the unprocessed speech $x_{in}$ and its style embedding vector $V=E_{s}(x_{in})$, along with the filtered speech $x^{\mathcal{M}}_{in}$.
We use a sensitivity score to measure the efficiency of flipping each dimension in $V$:
\begin{equation}
     Sen(x_{out}, x^{\mathcal{M}}_{in}, x_{in}) = \frac{\mathcal{L}_{spk}(x_{out}, x_{in}) - \mathcal{L}_{spk}(x^{\mathcal{M}}_{in}, x_{in})}{\Vert \mathbf{X}_{in}-\mathbf{X}_{out} \Vert_{2}},
\end{equation}
where $\mathbf{X}_{in}$ and $\mathbf{X}_{out}$ are spectrograms of $x_{in}$ and $x_{out}$.
Algorithm~\ref{Algorithm: audio style} presents the audio style optimization process. 
\begin{algorithm}[htbp]
    \caption{Style Embedding Optimization}
    \label{Algorithm: audio style}
    \begin{algorithmic}[1]
        \STATE \textbf{Input:} style vector $V = [v_0, v_1, \ldots, v_n]$, $x_{in}$ and $x^{\mathcal{M}}_{in}$
        \FOR {each $v_i \in \mathbf{V}$}
            \STATE $V_{temp} \gets V$ with $v_i = -v_i$
            \STATE $x_{out} = S(x^{\mathcal{M}}_{in}, V_{temp})$
            \STATE $Score[i] = Sen(x_{out}, x^{\mathcal{M}}_{in}, x_{in})$
        \ENDFOR
        \STATE Sort $Score[i]$ in descending order
        \FOR {each $i$ in sorted $Score[i]$}
            \WHILE {$\Vert X_{in}-X_{out} \Vert_{2} < \tau$}
                \STATE $\mathbf{V}_{out} \gets \mathbf{V}$ with $v_i = -v_{i}$
            \ENDWHILE
        \ENDFOR
    \end{algorithmic}
\end{algorithm}

First, we attempt to flip each dimension in $V$ and record the sensitivity, which is the ratio of the extra voice embedding loss to the audio quality loss caused by style transfer.
Next, we flip these dimensions according to the sensitivity score from high to low.
Meanwhile, we set an audio distortion threshold $\tau$.
Once the audio quality loss is beyond this threshold, we will quit the loop to ensure the audio distortion is acceptable with $x_{out} = S(x^{\mathcal{M}}_{in}, V_{out})$.
Although sounds similar, $x_{out}$ causes completely different voice embedding vector from the raw speech $x_{in}$.
\vspace{-7pt}
\subsection{Reverberation Generation}
\subsubsection{RIR Selection}
In the final step, we enhance the protection effectiveness by adding additional reverberation to the protected speech.
As mentioned in Section~\ref{Sec: Reverb}, reverberation is generated by convolving a reversed RIR with the raw audio signal. 
The selection of an appropriate RIR is guided by two main factors. 
On one hand, RIR signals are typically characterized as damped oscillatory signals with a rapid initial decay. 
To maintain the naturalness of the processed audio, the reverberation follows this characteristic. 
On the other hand, while reverberation is ubiquitous in physical environments, excessive reverberation can still degrade audio quality. 
Therefore, we must impose constraints on RIR seeds for reverberation generation.

First, the length of the RIR seed determines the perceptibility of the reverberation.
Typically, when the sound reflection delay is shorter than 30 ms, the reflected sound is mixed with the original sound~\cite{haas1972influence} so that it is imperceptible.
When the delay is longer, human ears identify the reflected sound as an echo, causing a significant difference in human perception.
Therefore, \ours constrains all RIR seeds length within 30 ms to guarantee audio clarity.
Second, to ensure naturalness, we collect RIR seeds from reverberation audios recorded in various real-world environments~\cite{traer2016statistics}.
We normalize all RIR samples and initialize the RIR seeds by clipping only the initial part of the RIR signal with relatively strong power.
Next, we attempt to select the RIR with the strongest protection performance, $h^*$, to maximize the loss of voice embedding vector, as shown in Eq.~(\ref{equation: speaker encoder}):
\begin{equation}
h^* = arg\max_{h \in S^{RIR}} ~ [\mathcal{L}_{spk}(x_{out} \ast h, x_{in})-\lambda_{l}\cdot len(h)],
\end{equation}
where $h$ represents the reversed RIR signals in the dataset $S^{RIR}$.
Meanwhile, we attempt to shorten the RIR length by incorporating a penalty term, $-\lambda_{l} \cdot len(h)$, to punish excessively long reverberation in the protected speech audio.
\subsubsection{Reverberation Optimization}
Natural reverberation alone may not always provide sufficient protection against malicious voice synthesis.
Therefore, the natural RIR seeds require further optimization to achieve the highest protection performance.
At this stage, the loss applied for optimization is different.
The previous goal of \ours optimization stages is to maximize the loss between the real voice embedding and the protected voice embedding.
However, continuously increasing the loss value does not always lead to better protection.
When the loss exceeds a certain threshold, the synthesized speech becomes noisy rather than merely altering the vocal features. 
This effect may alert adversaries to the presence of protective measures.
Our solution is to select a "target speaker" with a completely different voice from the protected speaker to effectively mislead the voice encoder.
AntiFake~\cite{yu2023antifake} applies a human-in-the-loop method, which asks human operators to select speech samples that are most dissimilar to the protected voice to constrain the voice embedding loss.
However, this method is labor-intensive, particularly in scenarios requiring large-scale protection.
Our approach addresses this limitation by employing a speaker recognition model to extract voice features and automatically select samples that are most dissimilar to the protected voice.
After it, the maximization problem is converted to a minimization problem.
The optimization can then be formulated as:
\begin{equation}
\begin{split}
\mathop{\textnormal{min}}\limits_{\delta}~[\mathcal{L}_{spk}(x^{tgt}, x_{out} \ast (h^* + \delta)) &- \lambda \cdot \mathcal{L}_{spk}(x_{in}, x_{out} \ast (h^* + \delta))]\\
\textbf{s.t.} \quad \Vert &\delta \Vert_{\infty} < \epsilon,
\end{split}
\label{Equation: RIR opt}
\end{equation}
where $x^{tgt}$ represents the speech sample from the target speaker, and $\lambda$ adjusts the weight of loss values.
Additionally, by introducing $\epsilon$, we can constrain the RIR amplitude and mitigate the interference caused by reverberation.

In addition, we optimize the projected gradient descent (PGD) method to enhance protection transferability.
We notice that, at the beginning of PGD, the adversarial examples show good transferability across different models.
However, as the PGD optimization continues, transferability tends to degrade.
After running PGD for multiple iterations, for $\mathcal{L}_{spk}$ in Eq.~(\ref{equation: speaker encoder}), the loss values of some encoders continue to decrease, while the loss values from other voice encoders remain unchanged or even increase.
To avoid this transferability loss, we implement a new strategy: for each individual voice encoder in the ensemble, if its voice embedding loss does not decrease over $k_c$ consecutive iterations, we stop the iteration.
This prevents the optimization process from overly focusing on a single model, which reduces overall protection transferability.
Furthermore, because our approach incorporates three different stages and a surrogate model composed of multiple encoders, it effectively enhances protection robustness. 
No voice synthesis model can withstand all of the protective methods or encoders.
    

\subsection{\ourss Design}~\label{section: livemask design}

While we can effectively protect offline speech , online streaming speech, such as online meetings or voice messages, is still vulnerable.
Compared to offline protection, online protection requires minimal latency, making step-by-step optimizations infeasible.
To address this challenge, we propose a fast protection model of \ours, named \ourss, to extend its feasibility for online applications.
In online protection, we skip audio style transfer because it requires the entire speech sample before processing.
In contrast, spectrogram masking and reverberation can be rapidly applied to real-time audio signals, enabling fast protection.

\revv{
Prior to the optimization, we prepare a dataset \( \mathcal{D} = \{\boldsymbol{x}_{0}, \boldsymbol{x}_{1},..., \boldsymbol{x}_{n} \} \), which contains 150 seconds speech samples from a single user, covering common English phonemes. 
}
In the first step, we optimize a general frequency mask $\mathcal{M}_{g}$ using the following objective:
\begin{equation}
arg \mathop{\textnormal{max}}\limits_{\mathcal{M}_{g}} ~ \sum^{n}_{i=0} \Vert Mel[\mathcal{M}_{g}(\boldsymbol{x}_i)] - Mel(\boldsymbol{x}_i) \Vert _2,
\end{equation}
where $\mathcal{M}_{g}$ is a filter that masks $k$ fixed frequencies ${f_0, f_1, ..., f_{k-1}}$ to maximize the mel-spectrogram loss across all samples in $\mathcal{D}$.
This mechanism pre-configures the filtered frequencies for the speaker.
In this way, when the microphone captures the streaming speech, the frequency filter can be applied in real-time without introducing additional latency.

Moreover, we design an optimization process for a universal RIR seed $h_g$ to generate reverberation in the speech.
Similar to the RIR optimization in \ours, this universal RIR seed $h_g$ is designed to minimize $\mathcal{L}_{spk}$ across all samples in $\mathcal{D}$.
The optimization process is formulated as:
\begin{equation}
\begin{split}
arg\mathop{\textnormal{min}}\limits_{\delta_{g}} ~ \sum_{i=0}^{n}[&\mathcal{L}_{spk}(x^{tgt}, x'_{i} \ast (h + \delta_g)) - \mathcal{L}_{spk}(x_{i}, x'_{i} \ast (h + \delta_g))] \\
&\textbf{s.t.} \quad \Vert \delta_g \Vert_{\infty} < \epsilon,~\textnormal{and}~~ h_g = h +\delta_g,
\end{split}
\label{Equation: fast rir opt}
\end{equation}
where $\boldsymbol{x}'_i$ represents the speech samples after universal frequency masking $\mathcal{M}_g$.
It is notable that finding the optimal solution for \ourss is more critical, as the universal RIR seed should indiscriminately protect ambient speech containing different contents from the given speaker.
To achieve this, we decrease the learning rate and increase the number of iterations to ensure the model finds a solution with better generalized performance.
Once the universal RIR seed is determined, the reverberation can be immediately applied to streaming speech signals via convolution.
Finally, the latency corresponds to the length of the RIR, which is typically tens of milliseconds.
Also, as we bypass the optimization process for individual speech samples, the masked universal filtered frequencies $k$ and the $\delta_g$ constraint $\epsilon$ in \ourss are less strict compared to the offline mode of \ours.

\vspace{-5pt}
\section{Evaluation}\label{sec:evaluation}
\subsection{Experiment Setup}
\begin{table}
\centering
\captionsetup{aboveskip=0pt, belowskip=0pt} 
\small
\caption{Voice synthesis models in \ours.}
\label{Table: models}
\begin{tabular}{|c|c|c|c|}
\hline
 & \begin{tabular}[c]{@{}c@{}}Voice Synthesis\\~Models\end{tabular} & \begin{tabular}[c]{@{}c@{}}Encoder\\~Architecture\end{tabular} & \begin{tabular}[c]{@{}c@{}}Embedding\\~Dimension\end{tabular} \\
\hline
\multirow{3}{*}{\begin{tabular}[c]{@{}c@{}}Surrogate\\models\end{tabular}} & AdaIN-VC~\cite{chou2019one} & VAE & 128 \\
\cline{2-4}
 & AutoVC~\cite{qian2019autovc} & VAE & 256 \\
\cline{2-4}
 & SV2TTS~\cite{wan2018generalized} & LSTM & 256 \\
\hline
\multirow{5}{*}{\begin{tabular}[c]{@{}c@{}}Test\\models\end{tabular}} & YourTTS~\cite{casanova2022yourtts} & ResNet & 512 \\
\cline{2-4}
 & DiffVC~\cite{popov2021diffusion} & VAE & 256 \\
\cline{2-4}
 & AGAIN-VC~\cite{chen2021again} & U-Net & N/A \\
 \cline{2-4}
 & ElevenLabs~\cite{elevenlabs} & N/A & N/A \\
 \cline{2-4}
 & Play.ht~\cite{playht} & N/A & N/A \\
\hline
\end{tabular}
\end{table}
\subsubsection{Speech Datasets}
We use two datasets for evaluation: one is VCTK-Corpus~\cite{veaux2016superseded}, an English dataset that contains 107 speakers with different accents.
Meanwhile, since most voice synthesis models are originally trained on this dataset, which may introduce bias into the results, we also use LibriSpeech~\cite{7178964}, another English speech dataset with longer speech samples.
We randomly select 50 speakers from each dataset (100 speakers in total) and test the 20 longest speech samples from each speaker.

\begin{table*}[htbp]
\centering
\captionsetup{aboveskip=3pt, belowskip=-5pt} 
\small
\caption{\rev{\ours performance on unseen open-source voice synthesis models.}}
\label{Table: Experimental results}
\begin{tabular}{|c|c|c|c|c|c|c|c|c|c|c|} 
\hline
\multirow{2}{*}{}                                                          & \multirow{2}{*}{Defenses}                                                                    & \multicolumn{3}{c|}{YourTTS}                                 & \multicolumn{3}{c|}{DiffVC}                                 & \multicolumn{3}{c|}{AGAIN-VC}                               \\ 
\cline{3-11}&                                                                                              & Score~$\downarrow$      & ETRR~ $\uparrow$ & SRR~ $\uparrow$ & Score~$\downarrow$      & ETRR~ $\uparrow$ & SRR~$\uparrow$ & Score~$\downarrow$      & ETRR~$\uparrow$ & SRR~$\uparrow$  \\ 
\hline
\multirow{8}{*}{\begin{tabular}[c]{@{}c@{}}Ablation\\Study\end{tabular}}   & N/A                                                                                          & 0.577$\pm$0.19          & 4.2\%            & 26.4\%          & 0.532$\pm$0.22          & 6.6\%            & 29.0\%         & 0.366$\pm$0.16          & 11.3\%          & 44.0\%          \\ 
\cline{2-11}& \ding{172}                                                                  & 0.202$\pm$0.09          & 75.6\%           & 91.2\%          & 0.209$\pm$0.09          & 71.3\%           & 90.0\%         & 0.177$\pm$0.07          & 84.0\%          & 96.5\%          \\ 
\cline{2-11}& \ding{173}                                                                  & 0.226$\pm$0.12          & 68.0\%           & 90.5\%          & 0.211$\pm$0.10          & 70.6\%           & 92.9\%         & 0.189$\pm$0.11          & 82.2\%          & 97.0\%          \\ 
\cline{2-11}& \ding{174}                                                                  & 0.187$\pm$0.08          & 83.5\%           & 94.5\%          & 0.188$\pm$0.11          & 88.4\%           & 96.9\%         & 0.158$\pm$0.13          & 95.0\%          & 99.1\%          \\ 
\cline{2-11}& \ding{172}+\ding{173}                                      & 0.176$\pm$0.09          & 94.2\%           & 100\%           & 0.164$\pm$0.08          & 95.3\%           & 100\%          & 0.120$\pm$0.06          & 100\%           & 100\%           \\ 
\cline{2-11}  & \ding{172}+\ding{174}                                      & 0.159$\pm$0.07          & 98.4\%           & 100\%           & 0.147$\pm$0.06          & 98.0\%           & 100\%          & 0.109$\pm$0.06          & 100\%           & 100\%           \\ 
\cline{2-11}  & \ding{173}+\ding{174}                                      & 0.163$\pm$0.07          & 96.5\%           & 100\%           & 0.152$\pm$0.07          & 98.8\%           & 100\%          & 0.114$\pm$0.05          & 100\%           & 100\%           \\ 
\cline{2-11}  & \textbf{\ding{172}+\ding{173}+\ding{174} }& \textbf{0.125$\pm$0.05} & \textbf{99.8\%}  & \textbf{100\%}  & 0.112$\pm$0.04 & 99.9\%  & \textbf{100\%} & 0.091$\pm$0.04 & \textbf{100\%}  & \textbf{100\%}  \\ 
\hline
\multirow{3}{*}{\begin{tabular}[c]{@{}c@{}}Existing\\Defenses\end{tabular}} & Attack-VC~\cite{huang2021defending}                                                                                    & 0.233$\pm$0.18              &    64.2\%              &   78.0\%              & 0.227$\pm$0.17              &  76.3\%               &  84.1\%              & 0.194$\pm$0.12                   &  82.5\%               &  93.9\%               \\ 
\cline{2-11}  & SampleMask~\cite{liu2023protecting}                                                                                   & 0.286$\pm$0.22              &    46.6\%              &  58.4\%               & 0.266$\pm$0.20              &   53.2\%               &  69.0\%              & 0.210$\pm$0.16             &    70.5\%             &    86.8\%             \\ 
\cline{2-11}  & AntiFake~\cite{yu2023antifake}                                                                                     & 0.138$\pm$0.06              &         \textbf{99.8\%}         &                 \textbf{100\%}& \textbf{0.107$\pm$0.04 }             &      \textbf{100\%}            &\textbf{100\%}                & \textbf{0.085$\pm$0.04 }                  &      \textbf{ 100\%}          &\textbf{100\%}                 \\
\hline

\multicolumn{11}{c}{\ding{172}~Spectrogram Masking ~~ \ding{173}~Audio Style Transfer~~\ding{174}~Reverberation Generation}

\end{tabular}
\end{table*}


\subsubsection{Speech Contents}
We use ChatGPT 4.0~\cite{ChatGPT} to generate 20 textual inputs that could potentially be used in voice deepfake attacks for TTS-based synthesis.
For VC-based synthesis methods, we use Google Text-to-Speech~\cite{GoogleTTS} to generate speech from these text inputs as the source speech.
Some textual input samples are included in Appendix~\ref{appendix_samples}.
\subsubsection{Voice Synthesis Models}
We employ multiple voice synthesis models for both the training and testing stages.
For surrogate models, we utilize AdaIN-VC~\cite{chou2019one}, AutoVC~\cite{qian2019autovc}, and SV2TTS~\cite{wan2018generalized}, which are based on Variational Autoencoder (VAE) and Long Short-Term Memory (LSTM).
We use 3 different open-source voice synthesis models for testing. 
YourTTS~\cite{casanova2022yourtts} is a TTS model with a ResNet-based voice encoder.
DiffVC~\cite{popov2021diffusion} is a voice conversion model using diffusion model to generate speech mel-spectrogram.
AGAIN-VC~\cite{chen2021again} is a voice conversion model using U-Net architecture to extract the content along with voice embedding vectors.
The details of these models are summarized in Table~\ref{Table: models}, where N/A indicates parameters that are either unknown or not applicable.
Although the basic architectures of these models are similar, their detailed parameters, such as the number of layers and weights, differ significantly, requiring \ours protection to achieve effective transferability.
Additionally, we test \ours performance on two commercial voice synthesis APIs: ElevenLabs~\cite{elevenlabs} and Play.ht~\cite{playht}.
These commercial APIs operate as pure black boxes, meaning we have no knowledge about their model architectures or weights.
\subsubsection{Hyperparameters}
We resample all speech audio sampling rate to 48 kHz in the pre-processing stage.
In spectrogram masking, we use STFT to transfer the audio waveform to a spectrogram with 1025 frequency bins mask $k$=12 frequencies.

\begin{table*}[htbp]
\centering
\captionsetup{aboveskip=3pt, belowskip=-5pt} 
\caption{\rev{\ourss performance on unseen open-source voice synthesis models.}}
\label{Table: livemask  results}
\small
\begin{tabular}{|c|c|c|c|c|c|c|c|c|c|} 
\hline
\multirow{2}{*}{Defenses} & \multicolumn{3}{c|}{YourTTS}                                & \multicolumn{3}{c|}{DiffVC}                                & \multicolumn{3}{c|}{AGAIN-VC}                               \\ 
\cline{2-10}
                          & Score~$\downarrow$      & ETRR~$\uparrow$ & SRR~ $\uparrow$ & Score~$\downarrow$      & ETRR~$\uparrow$ & SRR~$\uparrow$ & Score~$\downarrow$      & ETRR~$\uparrow$ & SRR~$\uparrow$  \\ 
\hline
\ding{172}                       & 0.211$\pm$0.16          & 70.2\%          & 88.5\%          & 0.221$\pm$0.18          & 62.6\%          & 80.4\%         & 0.188$\pm$0.13          & 79.4\%          & 91.7\%          \\ 
\hline
\ding{173}                       & 0.182$\pm$0.12          & 83.5\%          & 94.4\%          & 0.169$\pm$ 0.12         & 89.4\%          & 97.5\%         & 0.155$\pm$0.10          & 93.6\%          & 100\%           \\ 
\hline
\textbf{\ding{172}+\ding{173}}          & \textbf{0.154$\pm$0.08} & \textbf{99.7\%} & \textbf{100\%}  & \textbf{0.145$\pm$0.07} & \textbf{99.9\%} & \textbf{100\%} & \textbf{0.126$\pm$0.05} & \textbf{100\%}  & \textbf{100\%}  \\ 
\hline
VSMask~\cite{wang2023vsmask}                    &  0.226$\pm$0.17                       &    70.8\%             &    84.6\%             &           0.239$\pm$0.20              &  66.0\%               &    79.5\%            &    0.181$\pm$0.14                     &    85.8\%             &   95.2\%              \\
\hline

\multicolumn{10}{c}{\ding{172}~Universal Spectrogram Masking ~~ \ding{173}~Universal Reverberation Generation}

\end{tabular}
\end{table*}

\subsubsection{Evaluation Metrics}
We use three different metrics to comprehensively evaluate \ours performance on ASV:\\
\textbf{Similarity Score:} SpeechBrain~\cite{speechbrain} is an open-source speech toolkit built on PyTorch~\cite{paszke2019pytorch}.
Its speaker verification function is built on the state-of-the-art ECAPA-TDNN~\cite{desplanques2020ecapa} model.
The similarity score is calculated using the cosine similarity between the reference voice embedding and the inference voice embedding vectors.\\
\textbf{ECAPA-TDNN Rejection Rate (ETRR):} The ECAPA-TDNN model provides a speaker verification decision based on the similarity score.
If the similarity score falls below a certain threshold (default = 0.25), the voice is rejected by ASV.\\
\textbf{Soniox Rejection Rate (SRR):} We also evaluate \ours performance using Soniox~\cite{soniox}, a speaker identification service API.
We first enroll the original speech samples as the reference and then upload the synthesized speech.
If the synthesized voice is identified as an "unknown speaker," we consider \ours to have successfully prevented the voice deepfake attack.

\vspace{-5pt}
\subsection{\ours Effectiveness on Open-source Voice Synthesis Models}
First, we evaluate the performance of \ours on unseen voice synthesis models. 
To test the synthesis capabilities of the models, we use unprotected speech samples as reference inputs for open-source voice synthesis models and evaluate the similarity between the synthesized and original voices using the ECAPA-TDNN model. 
The results are listed in the first row of Table~\ref{Table: Experimental results}. 
Although the synthesis performance varies across different models, most synthetic voice samples are successfully verified by the ECAPA-TDNN speaker verification system across all models. 
This demonstrates that sharing unprotected speech online can easily compromise ASV systems.

Next, we conduct an ablation study by comparing each step in \ours individually and in different combinations.
When we apply only the spectrogram masking method, the similarity between the synthesized voice samples and the unprotected synthesized voices significantly decreases across all models, indicating good transferability of this approach. 
However, despite the reduction in similarity, many synthesized voice samples can still pass speaker verification as voice features still remain in the residual frequencies. 
Moreover, further increasing the number of masked frequencies would result in a substantial decline in audio quality. 
Therefore, simply relying on this method is insufficient to comprehensively protect human speech against voice synthesis. 
In addition, we evaluate the protective performance of the audio style transfer and reverberation generation methods. 
Compared to spectrogram masking, these two methods are optimized based on the loss provided by the surrogate ensemble encoders. 
In our experiments, both methods demonstrate high protective effectiveness. 
Despite significant differences in structure and parameters across different models, the aggregate encoder approach we use has a noticeable impact on all models. 
After applying audio style transfer, most synthesized voice samples can no longer bypass speaker verification. 
Similarly, the reverberation generation method successfully protects over 80\% of the speech samples. 
However, when used individually, these methods can still be compromised by attackers by repeated attempts.

For comparison, we run experiments with different combinations of these three methods. 
As shown in Table~\ref{Table: Experimental results}, the protective effectiveness for unseen models significantly increases when multiple methods are applied.
When two of the methods are used in combination, at least 94\% of the voice samples are successfully protected.
When all three methods are combined, nearly all samples are successfully protected across all potential attack models.
Considering that the ECAPA-TDNN model has an error rate of approximately 1\%, \ours remains effective even if the rejection rate is not always 100\%\footnote{Audio demos are available at:~\hyperref[https://clear-mask.github.io/]{https://clear-mask.github.io/}.}.

Meanwhile, we evaluate the effectiveness of \ours on Soniox, a commercial speaker recognition API.
Generally, commercial ASV models have more complex architectures, leading to better robustness than open-source models.
Since commercial APIs will be used in real-world application scenarios to process diverse speech data, the ASV model must ensure the lowest possible false acceptance rate (FAR), which means that it has a higher threshold to reject deepfake or similar voice samples.
Even generated from unprotected samples, a lot of synthetic voices fail to pass the commercial ASV models.
When we apply different protection combinations, most methods successfully achieve a 100\% protection success rate.
Therefore, \ours demonstrates strong effectiveness in preventing all synthetic speech samples from passing commercial ASV models.

\rev{
In addition, we compare the performance of \ours with other baseline offline defense methods, including Attack-VC, SampleMask, and AntiFake. 
For Attack-VC and SampleMask, we adopt the white-box defense setup. 
We choose the samples with the best transferability performance across multiple target voice synthesis models. 
As shown in Table~\ref{Table: Experimental results}, when Attack-VC and SampleMask are applied to defend against unseen voice synthesis models, they fail to achieve effective results due to their inability to handle varying model structures and parameters.
After applying these defenses, the synthesized voices still exhibit high similarity to the original voices, failing to meet the transferability requirements for defending against deepfake voices.
In comparison, AntiFake leverages an ensemble encoder approach to improve transferability, and our experiments confirm its effectiveness. 
Similar to \ours, AntiFake achieves nearly 100\% defense success rates in black-box scenarios. 
However, AntiFake suffers from significant degradation in the quality of the protected audio, which we will compare in detail in Section~\ref{sec: quality_comparison}.
}

\subsection{\ourss Effectiveness Evaluation}
Next, we apply \ourss protected speech to the same open-source voice synthesis models.
As introduced in Section~\ref{section: livemask design}, \ourss does not involve optimization for individual speech samples.
Therefore, we modify the constraint hyperparameters used in \ours.
In the universal spectrogram masking stage, we mask $k = 16$ fixed frequencies.
In the reverberation generation stage, the length of $h_g$ is fixed to 30 ms.

Similar to offline \ours evaluation, we assess the performance of each stage and \ourss with both stages in the ablation study.
The experimental results are shown in Table~\ref{Table: livemask results}.
In the spectrogram masking step, the universal protection method only masks fixed frequencies.
However, removing more frequencies is not feasible given the audio quality requirements.
Additionally, due to the diversity of human languages, this fixed protection strategy cannot guarantee effectiveness for all speech content.
Consequently, we observe that while universal spectrogram masking reduces the similarity between synthesized and original speech compared to unprotected speech, as shown in Table~\ref{Table: Experimental results}, the average similarity still remains relatively high, and the standard deviation is noticeably greater than \ours.
Such result indicate that for some speech samples, only masking fixed frequencies is not sufficiently effective.
Similarly, the universal reverberation generator can also reduce the similarity between synthesized voices and the original voice, but it fails to provide complete protection.
In comparison, when we combine the two stages, the protection demonstrates better transferability and effectiveness across all models.
At least 99.7\% of cloned voices cannot bypass the ECAPA-TDNN ASV model, and none of the samples can spoof the Soniox ASV model.
Therefore, \ourss achieves effective protection across various unseen voice synthesis models.
\revv{We further assess the computational requirements of \ourss.
Specifically, when applying reverberation to a 48 kHz audio signal, the computational cost is 70 million floating-point operations per seconds (FLOPs), with a memory footprint of 80 MB. 
These results demonstrate that \ourss is capable of running on lightweight devices, such as the Raspberry Pi 4, which features at least 1 GB of random-access memory (RAM) and supports 3 billions FLOPs.
Moreover, the computational efficiency of \ourss can be further optimized for IoT devices by reducing the audio sampling rate, thereby lowering both processing and memory demands.}

\rev{
Moreover, we compare \ourss with the existing online protection method, VSMask. 
We find that VSMask is ineffective against unseen voice synthesis models. 
This is because its predictive model is trained in a white-box setup and heavily depends on the weights of a specific voice synthesis model. 
As a result, the generated perturbations fail to maintain their effectiveness across different models.
}

\subsection{\ours and \ourss Performance on Commercial Platforms}
Next, we evaluate the performance of \ours and \ourss on commercial voice cloning platforms.
Similar to ASV models, commercial voice cloning services typically use larger-scale models and more extensive training data than open-source voice synthesis models, resulting in greater robustness.
Moreover, these models are entirely black-box, meaning we can only counter them by enhancing the transferability of our protection methods.

Compared to open-source speech synthesis models, commercial models require longer speech samples to improve synthetic speech quality.
Therefore, we concatenate the protected samples into longer voice samples.
We compare the voice rejection rates of the ECAPA-TDNN and Soniox ASV models before and after applying \ours and \ourss protection.
The experimental results are presented in Table~\ref{Table: API results}.
According to the results, commercial voice cloning platforms achieve lower voice rejection rates than open-source voice synthesis models.
Almost all synthetic speech samples generated from unprotected speech can successfully spoof the ECAPA-TDNN model, and over 80\% can even bypass the Soniox speaker recognition API, revealing a severe threat to user privacy.
Additionally, because these commercial platforms do not require any local resources, such as GPUs or development environments, they present a greater threat than open-source models.
With just 10 seconds of unprotected clear speech, these commercial models can generate a large number of high-quality and highly similar synthetic speech samples.

Next, we upload voice samples protected by \ours to the API and test the similarity between the generated voices and the original voices.
Despite having no knowledge of the internal workings of the API models, \ours still demonstrates high transferability.
Over 98\% of the voice samples generated from \ours protected samples cannot pass the ECAPA-TDNN based ASV model, and 100\% are rejected by the commercial speaker recognition API.
Additionally, \ourss can also successfully protect over 97\% of samples in real-time.
As a result, both \ours and \ourss demonstrate strong protection effectiveness on black-box voice synthesis API platforms.
\begin{table}
\centering
\small
\caption{\ours and \ourss performance on commercial voice synthesis platforms.}
\label{Table: API results}
\begin{tabular}{|c|c|c|c|c|} 
\hline
\multirow{2}{*}{Method} & \multicolumn{2}{c|}{ElevenLabs}   & \multicolumn{2}{c|}{Play.ht}      \\ 
\cline{2-5}
                        & ETRR~$\uparrow$ & SRR~$\uparrow$  & ETRR~$\uparrow$ & SRR~$\uparrow$  \\ 
\hline
N/A                     & 1.0\%          & 19.3\%          & 0.0\%          & 12.2\%          \\ 
\hline
\ourss                        & 98.8\%          & 100\%          & 97.4\%          & 100\%          \\ 
\hline
\ours                      & \textbf{99.2\%} & \textbf{100\%}  & \textbf{98.3\%} & \textbf{100\%}  \\
\hline
\end{tabular}
\end{table}

\subsection{Human Perception Evaluation}
To test the performance of \ours for prevent synthetic speech from spoofing human perception, we use perceptual speaker dissimilarity (PSD) to measure it.
The PSD score ranges from 1 to 5, where 1 indicates that the two voices are from the same speaker, and 5 indicates that the two voices are distinct.
To save human effort, we select the best-performing open-source and commercial voice cloning models, YourTTS and Play.ht, as the adversarial models to verify the effectiveness of \ours and \ourss.
We collected over 550 responses from 28 listeners (17 males and 11 females, aged 20 to 35) with normal hearing ability, who are asked to rate the PSD between the two voice samples.

\begin{figure}[tbp]
    \centering
    \includegraphics[width=0.46\textwidth]{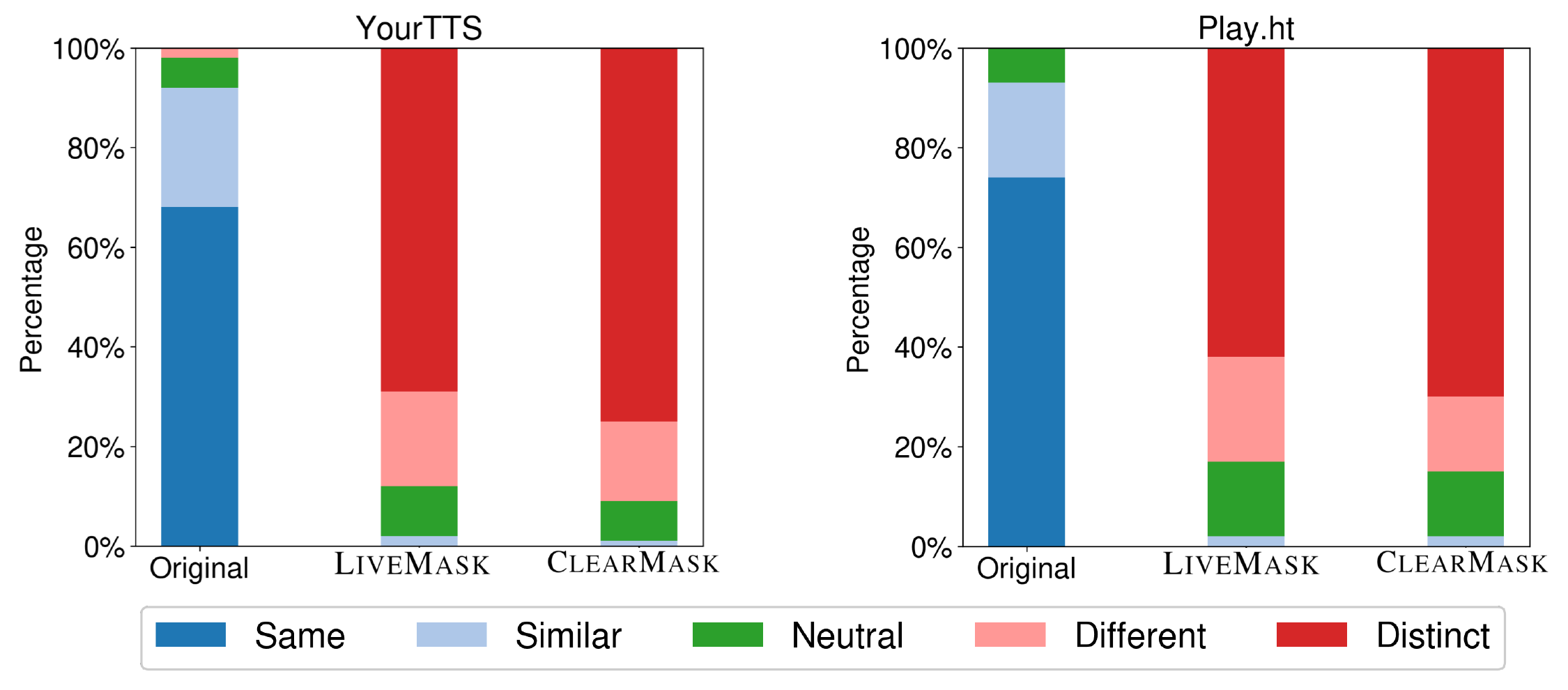}
    \caption{PSD comparison of synthetic voice generated from unprotected and protected speech samples.}
    \setlength{\belowcaptionskip}{-20pt}
    \vspace{-15pt}
    \label{fig: PSD}
\end{figure}

We present the PSD results in Fig.~\ref{fig: PSD}.
When the unprotected voice samples are exposed to attackers, the synthetic speech from advanced voice synthesis models can easily deceive human listeners.
Over 90\% of responses indicate that the synthetic voice is the same as or similar to the reference voice.
These synthetic speech samples are highly likely to fool listeners into misjudging the speaker's identity, potentially leading to financial losses or security threats.
In comparison, with \ours and \ourss protection, the synthetic voice sounds distinct from the real voice.
Less than 2\% of the samples are identified as "similar to" the real voice. 
None of the samples are considered to be from the same speaker as the reference voice.
When attackers use protected speech to synthesize deepfake voice, the resulting speech not only exhibits different characteristics from the original voice but also sounds hoarse and unstable.
As a result, the synthetic speech can be easily recognized by human listeners as not being from the target speaker, resulting in a failed attack. 
\subsection{\ours Audio Quality Comparison with AntiFake}\label{sec: quality_comparison}
One of the main innovations of \ours and \ourss is that the strategies we apply, including spectrogram masking, audio style transfer, and reverberation generation, do not introduce ambient noise to the original clear speech.
This allows \ours and \ourss to achieve better audio quality and naturalness compared to other existing voice deepfake defenses based on perturbation optimization.
In this section, we use different evaluation metrics to measure the audio quality of protected speech samples from various protection methods.\\
\textbf{Mean Opinion Score (MOS):} A measurement of audio quality based on subjective listener ratings.
We use NISQA, a DNN-based method, to estimate speech quality and naturalness on a scale from 1 to 5.
Typically, when the MOS is higher than 3, the speech is considered clear for human perception.\\
\textbf{Short-Time Objective Intelligibility (STOI):} STOI measures the intelligibility of the processed input signal by comparing it with the clean reference signal.
The STOI score ranges from 0 to 1, representing speech that is absolutely unintelligible to highly intelligible.\\
\textbf{Perceptual Evaluation of Audio Quality (PEAQ):} PEAQ is a subjective audio quality evaluation based on real human perception.
Listeners evaluate their perceived audio quality and provide an opinion score on a scale from 1 (very noisy) to 5 (imperceptible noise).

\begin{table}[tbp]
\centering
\small
\caption{\ours and \ourss protected audio quality comparison with AntiFake.}
\label{Table: Audio Quality}
\begin{tabular}{|c|c|c|c|} 
\hline
           & MOS~$\uparrow$ & STOI~$\uparrow$ & PEAQ~$\uparrow$  \\ 
\hline
AntiFake~\cite{yu2023antifake}   & 2.82$\pm$0.66    &  0.34$\pm$0.11 & 2.41$\pm$0.83        \\ 
\hline
\ourss      & 3.01$\pm$0.80    & 0.59$\pm$0.12     & 3.77$\pm$0.77\\        \hline
\ours       & \textbf{3.12$\pm$0.72}    & \textbf{0.65$\pm$0.09}     & \textbf{4.26$\pm$0.51}\\ 

\hline
\end{tabular}
\end{table}

\begin{figure}[t]
    \centering
    \captionsetup{aboveskip=0pt, belowskip=-10pt}
    \subfigure[The original speech.]{\includegraphics[width=0.20\textwidth]{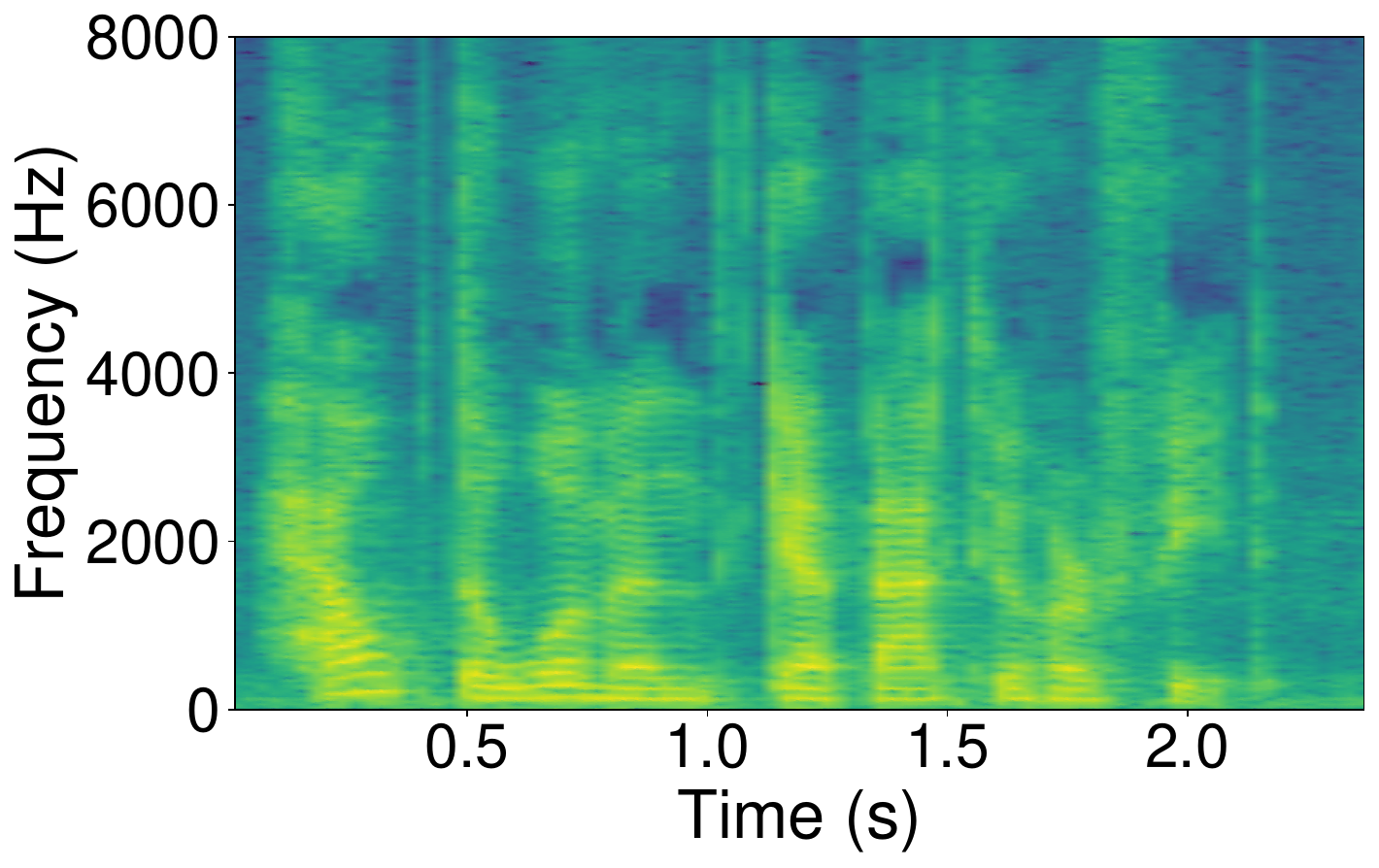}
    \label{fig: spec1}}
    \
    \subfigure[The speech protected by \ours.]{\includegraphics[width=0.20\textwidth]{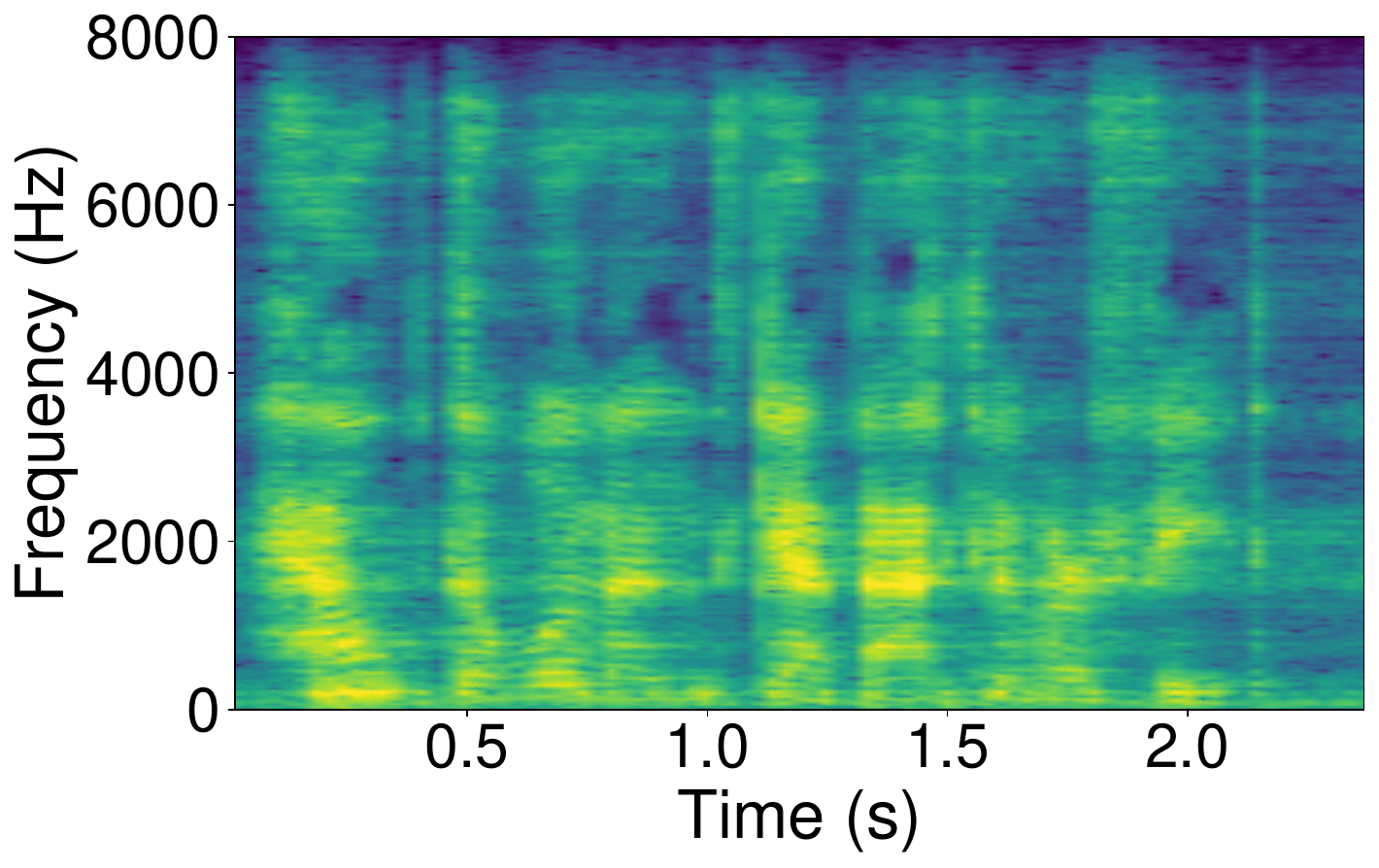}
    \label{fig: spec2}}   
    
    \subfigure[The speech protected by \ourss.]{\includegraphics[width=0.20\textwidth]{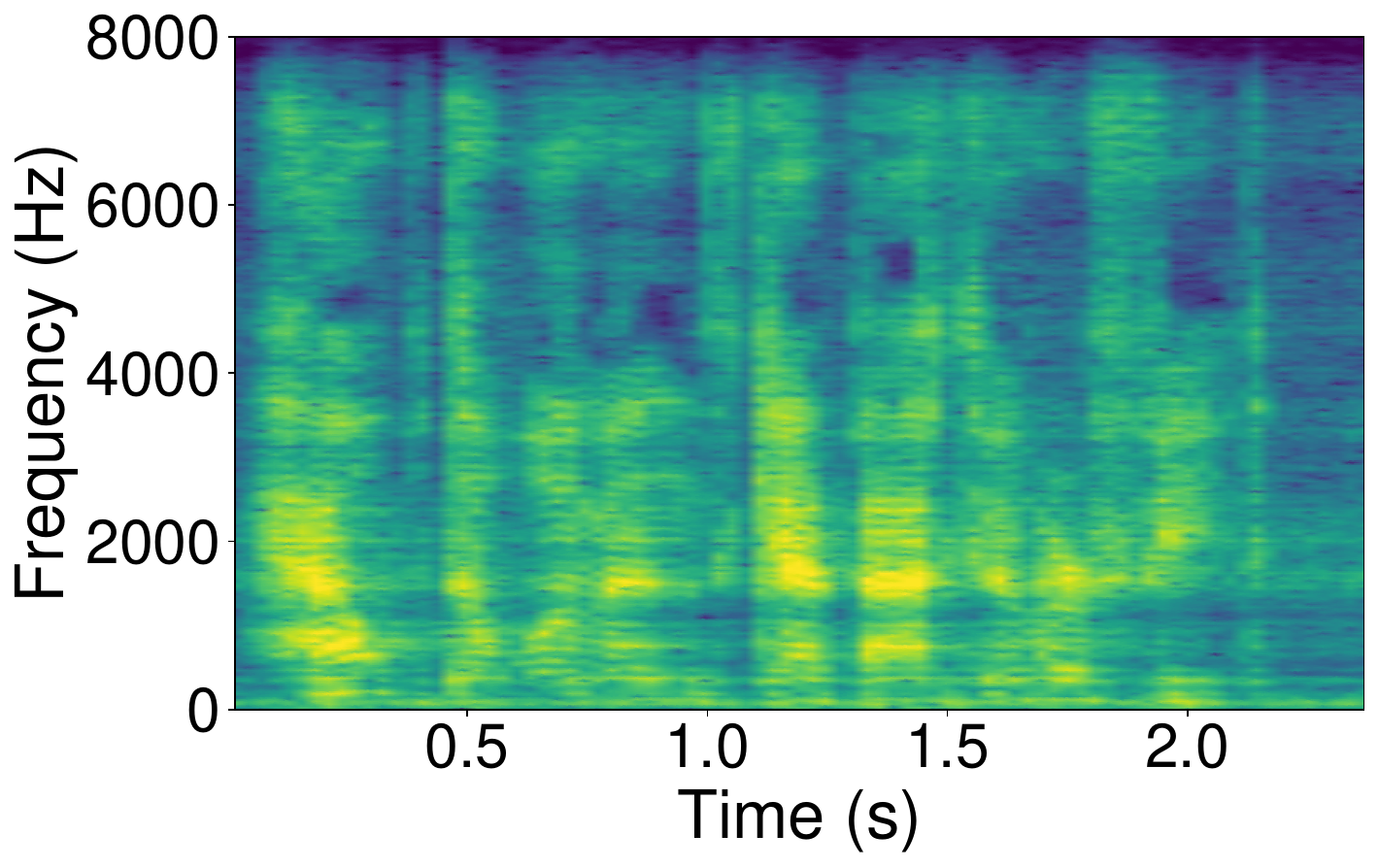}
    \label{fig: spec3}}
    \
    \subfigure[The speech protected by AntiFake.]{\includegraphics[width=0.20\textwidth]{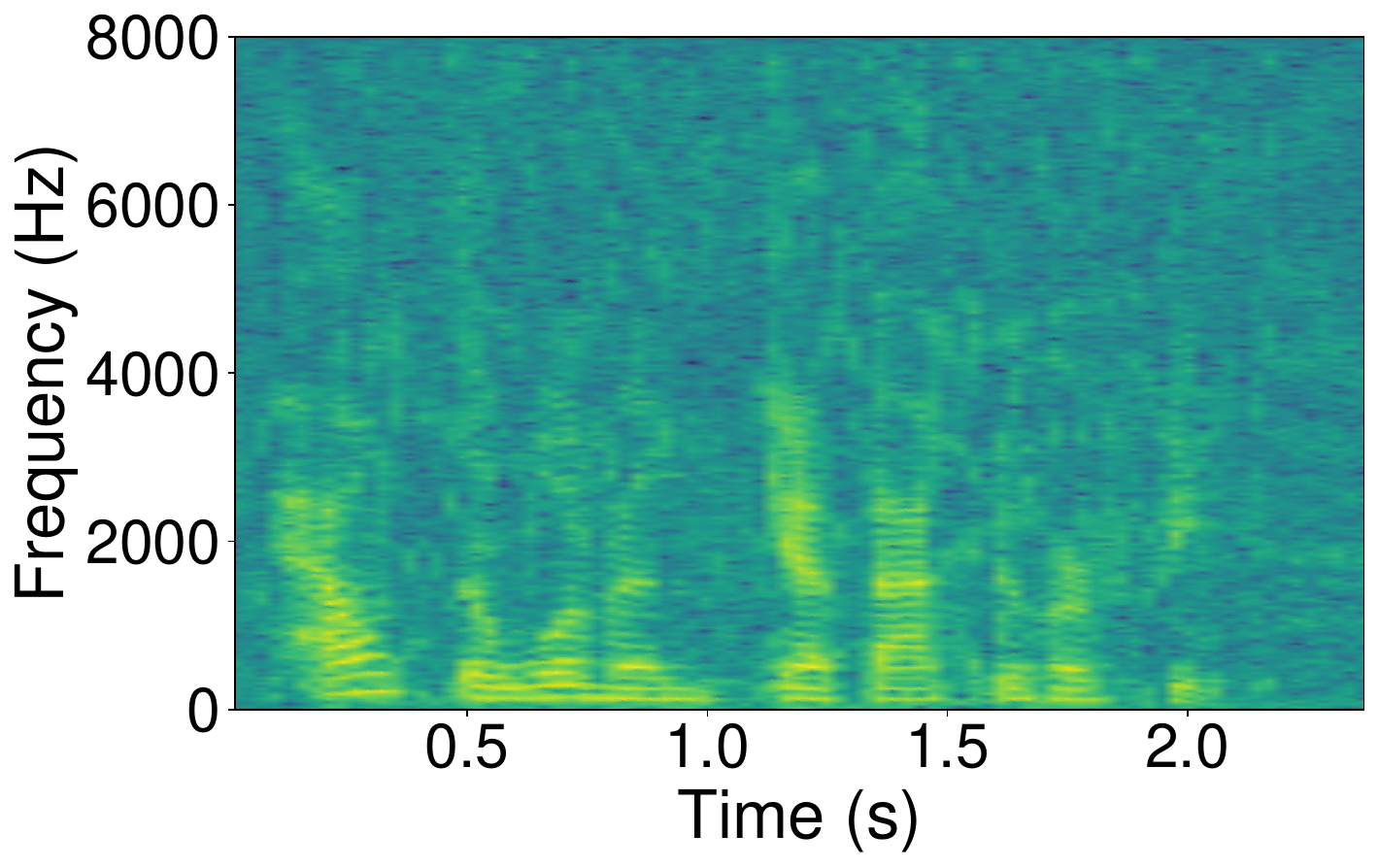}
    \label{fig: spec4}}
    \caption{Spectrograms of raw speech and protected speech samples from \ours, \ourss and AntiFake.}
    \vspace{-15pt}
    \label{fig: audio quality comparison}    
\end{figure}

Considering that white-box defense speech samples may have better audio quality, this trade-off compromises their ability to achieve transferable defense, making them less practical for use.
In the evaluation, we only compare AntiFake~\cite{yu2023antifake} with \ours and \ourss.
The comparison results are listed in Table~\ref{Table: Audio Quality}.
For MOS, \ours is slightly better than \ourss and the AntiFake methods.
Although \ours and \ourss do not apply noise injection, the MOS measurement is sensitive to audio "coloration," which refers to artificial patterns. 
Therefore, additional reverberation can also degrade the opinion score.
In contrast, \ours and \ourss show much better results in STOI and PEAQ.
In Fig.~\ref{fig: audio quality comparison}, we display spectrograms from unprotected speech sample along with the protected samples generated by \ours, \ourss, and AntiFake.
We observe that while \ours inevitably causes some loss in audio quality compared to unprotected samples, the protected audio still retains most of its original features.
As a result, \ours achieves relatively high STOI and PEAQ scores.
In contrast, \ourss introduces additional reverberation and signal loss compared to \ours, resulting in slightly lower scores.
By comparison, in AntiFake-protected audio, noise is dispersed across the entire frequency spectrum, with high-frequency signals above 4 kHz almost entirely overwhelmed by adversarial noise.
Although AntiFake attempts to mitigate the impact of noise based on human auditory sensitivity, it still severely degrades the quality and intelligibility of the sound, making it unsuitable for applications where high audio quality is required.

\begin{figure*}[htbp]
    \centering
    \begin{minipage}[t]{0.30\textwidth}
        \centering
        \includegraphics[width=\textwidth]{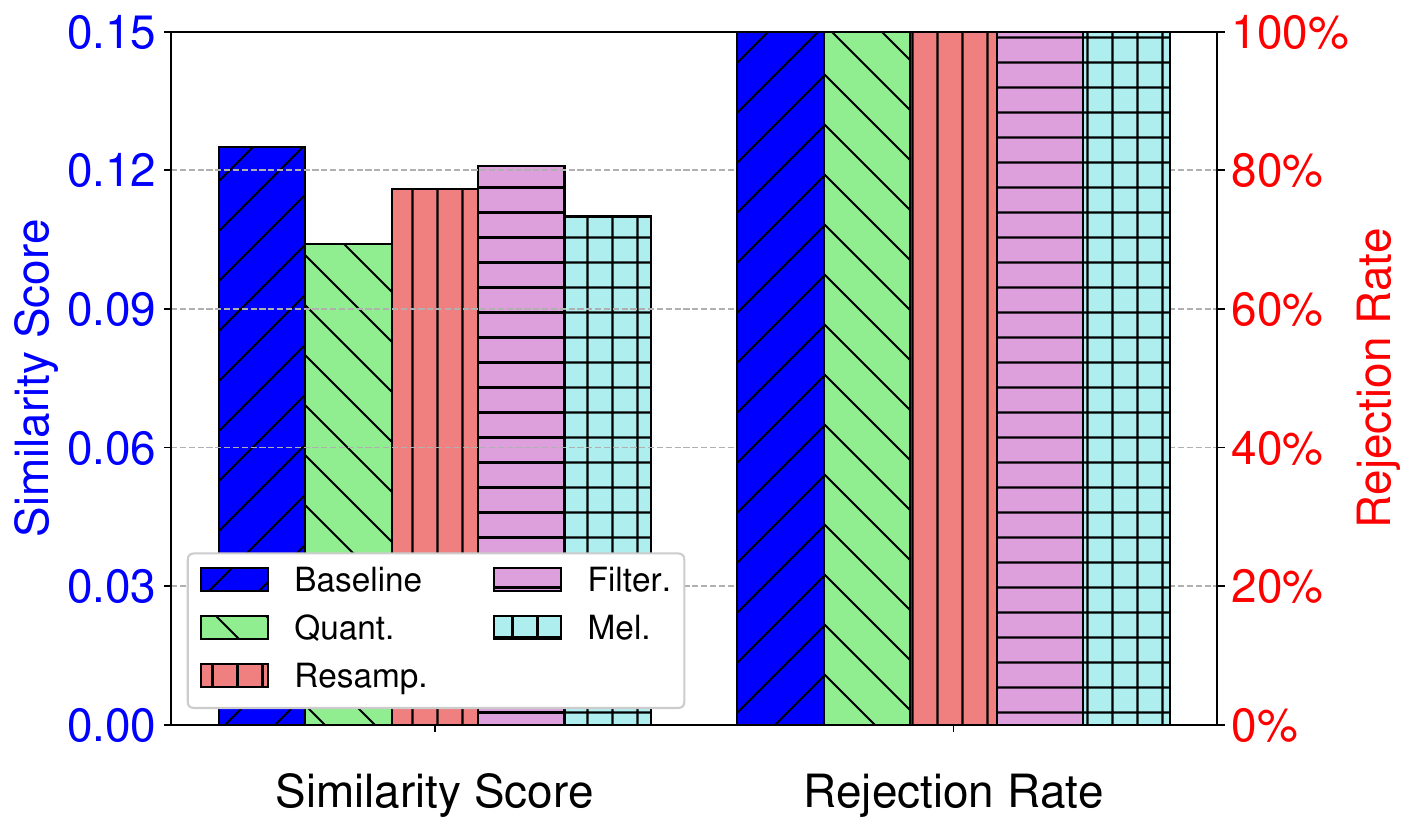}
        \vspace{-10pt}
        \caption{\rev{\ours performance when adaptive attackers apply WaveGuard to recover the speech.}}
        \label{fig:adaptive_r1}
    \end{minipage}
    \hfill
    \begin{minipage}[t]{0.30\textwidth}
        \centering
        \includegraphics[width=\textwidth]{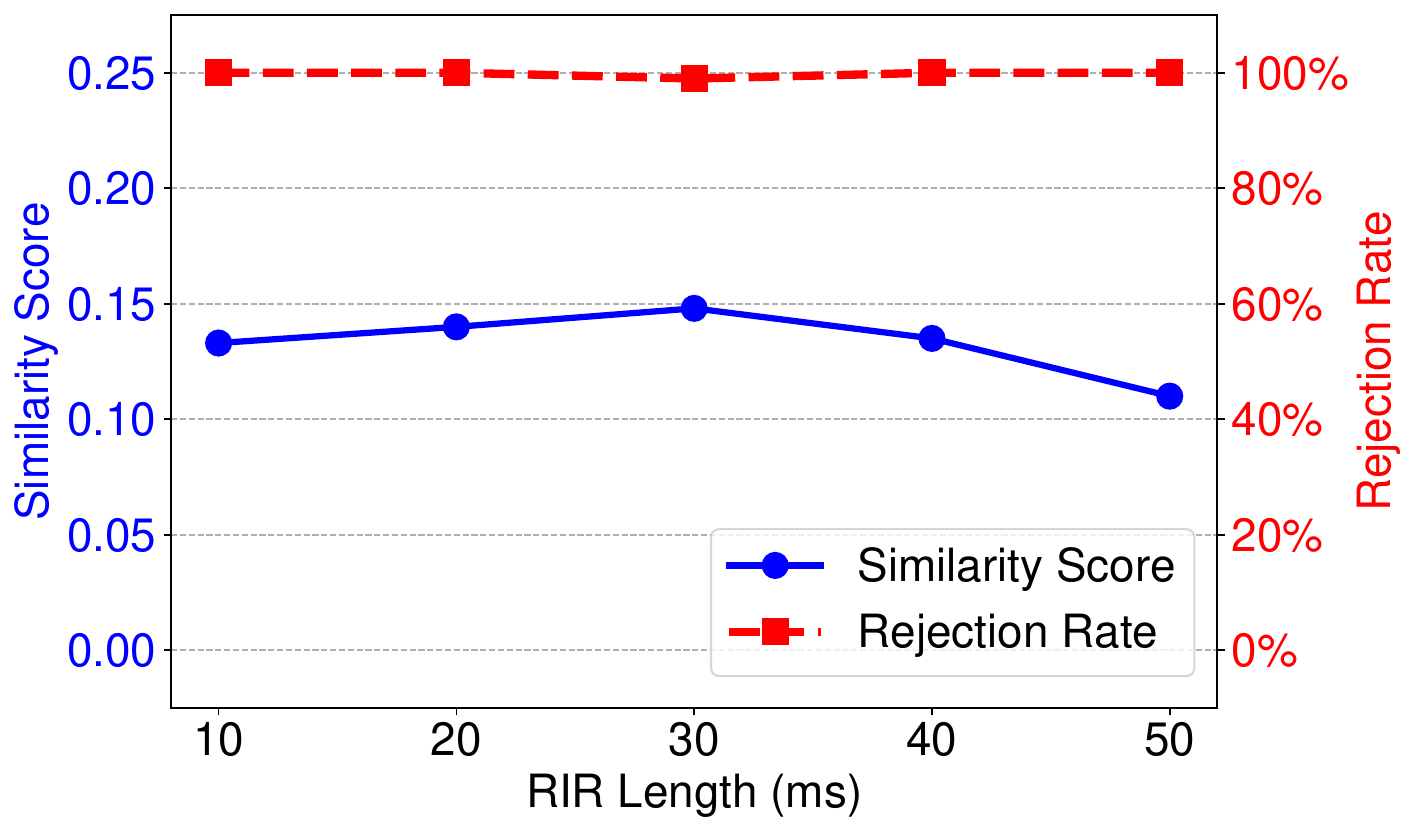}
        \vspace{-10pt}
        \caption{\rev{\ours performance against adaptive deconvolution-based reverberation removal.}}
        \label{fig:adaptive_r2}
    \end{minipage}
    \hfill
    \begin{minipage}[t]{0.30\textwidth}
        \centering
        \includegraphics[width=\textwidth]{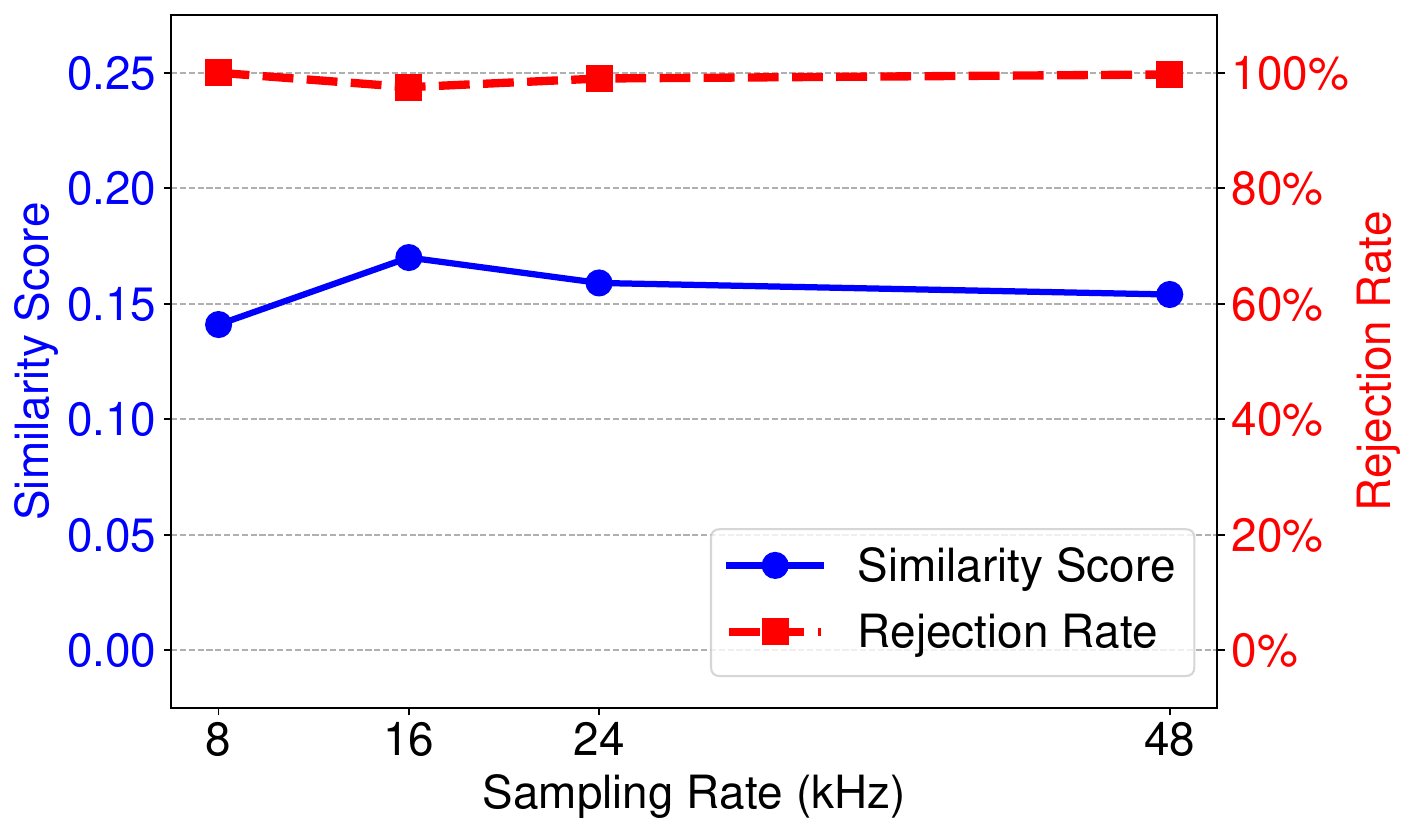}
        \vspace{-10pt}
        \caption{\rev{\ourss performance comparison under different audio sampling rates.}}
        \label{fig:livemask_sr}
    \end{minipage}
    \vspace{-10pt}
\end{figure*}
\subsection{Adaptive Attackers}

\rev{We consider two types of adaptive attackers. 
The first type, referred to as \textbf{R1}, has no prior knowledge and attempts to recover the original audio using conventional signal processing techniques, such as WaveGuard~\cite{hussain2021waveguard}. 
The second type, referred to as \textbf{R2}, is aware of \ours's defense mechanism and employs deconvolution methods to remove reverberation and restore the original audio.}

\rev{
\noindent \textbf{R1 Attacker:} 
We implement four transformations in WaveGuard: Quantization-Dequantization (Quant.), Down/Up-sampling (Resamp.), Frequency Filtering (Filter.), and Mel-spectrogram Inversion (Mel.).
Specifically, we quantize the audio to 8 bits and down-sample it to a frequency of 8 kHz.

The results are presented in Fig.~\ref{fig:adaptive_r1}. 
Compared to the raw protected audio (Baseline), WaveGuard-processed samples fail to recover the voice features.
None of these transformations enable synthetic speech to bypass the ASV model. 
This is because WaveGuard attempts to eliminate subtle perturbations by compressing the audio features.
However, these approaches are ineffective against the defense mechanisms in \ours, as the protected audio contains no noise-based perturbations to exploit.\\
\textbf{R2 Attacker:}
For more sophisticated attackers who have knowledge about \ours, they attempt to remove the adversarial reverberation effect in the protected speech.
In this experiment, the RIR length in \ours is fixed at 30 ms.
However, the length is an unknown hyperparameter for the adaptive attackers, so they would adopt the same ensemble encoder to optimize RIRs of varying lengths for deconvolution.

In Fig.~\ref{fig:adaptive_r2}, we show the impact of adaptive attackers using YourTTS for deepfake voice generation.
We find that regardless of the length of the natural RIR used by adaptive attackers for deconvolution to attempt to remove reverberation, they consistently fail to generate qualified cloned voices from the processed speech samples.
The reason is that both spectrogram masking and audio style transfer play critical roles in \ours protection.
Moreover, these techniques are irreversible, meaning that even with comprehensive prior knowledge, attackers cannot mitigate their effects.
Second, once the speech sample is protected with \ours, the voice embedding vector becomes similar to an unknown "target speaker" voice.
In this way, adaptive attackers are unaware of the real voice embedding and will attempt to push the voice embedding vector far from its initial state.
Thus, this attack approach will not succeed in recovering the genuine voice features.
}

\revv{Moreover, adversarial training can enhance the robustness of speech synthesis models against protected samples. To test its feasibility, we apply adversarial training to AdaIN-VC using adversarial samples generated via PGD with 1,000 iterations. 
However, this process requires 150 times more training time (over 20 days) than standard training, making it excessively costly and potentially impractical for attackers.
Additionally, it further degrades the model’s ability to synthesize ordinary speech~\cite{madry2018towards}.
Therefore, we do not discuss it in detail.}

\vspace{-5pt}
\section{Discussion}
\noindent\textbf{\ours Protection Performance in Other Languages.}
\rev{
In Section~\ref{sec:evaluation}, we evaluated the protection performance of \ours and \ourss.
However, since different languages exhibit distinct phonetic characteristics, it is essential to test the effectiveness of \ours and \ourss in other languages as well. 
In this experiment, we test \ours and \ourss in Mandarin and Spanish, two of the most widely spoken languages in the world besides English, to validate the transferability of \ours across multiple languages.
All datasets are collected from the Mozilla Common Voice~\cite{ardila2019common}.
We use YourTTS to generate deepfake voices, as its foundation model is trained on multiple languages to ensure that the speech synthesis capability remains consistent across different target languages.

We use average similarity score (ASS) and ETRR to measure the protection effectiveness. 
The experimental results are listed in Table~\ref{Table: languages}. 
When no protection is applied, the model's ability to synthesize Mandarin and Spanish speech is slightly weaker than English, due to differences in the amount of training data. 
Nevertheless, the synthesized speech still exhibits high voice similarity, with over 90\% of the voice samples successfully passing ECAPA-TDNN speaker verification.
In contrast, when we apply \ours and \ourss protection to the reference voice samples, the similarity of the synthesized Mandarin and Spanish voice samples is significantly reduced.
This demonstrates that our protection strategy is effective across different languages, even when their pronunciation and intonation are different.
}

\begin{table}
\centering
\small
\caption{\rev{\ours and \ourss protection performance in different languages.}}
\label{Table: languages}
\begin{tabular}{|c|c|c|c|c|c|c|} 
\hline
\multirow{2}{*}{Language} & \multicolumn{2}{c|}{Raw Speech} & \multicolumn{2}{c|}{\ours}          & \multicolumn{2}{c|}{\ourss}  \\ 
\cline{2-7}
                          & ASS~$\downarrow$   & ETRR~$\uparrow$         & ASS~$\downarrow$ & ETRR~$\uparrow$ & ASS~$\downarrow$   & ETRR~$\uparrow$                \\ 
\hline
\revv{English}                  & 0.577 & 4.2\%                   & 0.125            & 100\%           & 0.154 & 99.8\%              \\ 
\hline
Mandarin                  & 0.496 & 7.6\%                   & 0.119            & 100\%           & 0.151 & 99.6\%              \\ 
\hline
Spanish                   & 0.520 & 5.5\%                   & 0.128            & 100\%           & 0.159 & 99.0\%              \\
\hline
\end{tabular}
\end{table}

\rev{

\noindent \textbf{\ourss Performance under Different Sampling Rates.}
In real-world scenarios, the audio sampling rate for real-time voice communication is constrained by network speed and hardware limitations, including microphones. 
To assess the effectiveness of \ourss under different conditions, we evaluate its performance across various sampling rates.

As illustrated in Fig.~\ref{fig:livemask_sr}, we test audio sampling rates of 8 kHz, which is used in telephone communication, and higher rates including 16 kHz, 24 kHz, and 48 kHz. 
Next, we synthesize deepfake voices from resampled audios on YourTTS model.
It is important to note that we do not design distinct reverberations for different sampling rates. 
Instead, we down-sample the RIR according to the audio’s sampling rate.
From the results, we observe that \ourss achieves optimal defensive performance at the default 48 kHz sampling rate. When the audio is sampled at 16 kHz, the defensive performance has a slight decline but still maintains a protection success rate above 97\%. 
At 8 kHz, due to the significantly reduced sampling rate, the effectiveness of deepfake voice synthesis is inherently diminished.
Overall, these results demonstrate that \ours delivers robust protective performance across a range of hardware and network conditions, maintaining its effectiveness even under varying sampling rate constraints.
}

\noindent \textbf{\ours Latency.}
We also compare the latency of \ours with other defense methods. 
When the number of iterations is set to 1000, Attack-VC~\cite{huang2021defending} takes an average of 35 seconds to generate a protected sample. 
AntiFake~\cite{yu2023antifake} takes about three minutes to provide protection due to the computational cost of gradient calculations across multiple models. 
Additionally, its human-in-the-loop mechanism for voice similarity assessment adds further latency.
In contrast, while \ours also computes gradients across multiple models, it optimizes an RIR signal of under 1,440 samples at 48 kHz. 
Other methods modify the mel spectrogram, requiring gradient calculations over thousands of dimensions, making \ours significantly faster than AntiFake.
The average time for \ours reverberation generation is 40 seconds, and when combined with the computation time for spectrogram masking and audio style transfer, the total average time is 60 seconds. 
For \ourss, since frequency filtering and convolution can be performed in real time, we consider its latency to be the length of the RIR signal, which is approximately 30 milliseconds. 

\noindent \textbf{Ethical Statement.}
We are deeply committed to addressing potential ethical issues associated with this research. 
We exclusively utilize publicly available speech datasets and open-source speech synthesis or speaker verification models, following their usage guidelines and licenses. 
All human listeners have provided informed consent after being informed of the study's purpose and procedures before participation, and the experiments involving human subjects have received IRB approval. 
Their privacy and personal information are also rigorously protected throughout the research process.

\vspace{-5pt}
\section{Related Work}
\subsection{Voice Synthesis and Countermeasures}
Many VC and TTS models have been developed to embed a voice onto given linguistic content. 
Early voice conversion methods relied on parallel speech data for model training~\cite{chen2003voice}.
Recent VAE-based voice conversion approaches are able to learn from different speech contents from various speakers~\cite{chou2019one}.
AutoVC~\cite{qian2019autovc} introduces a bottleneck mechanism to enable zero-shot voice conversion.
Different from VC, TTS can generate speech only according to textual input.
SV2TTS~\cite{jia2018transfer} can embed ambient speaker's voice into the synthetic speech.
Currently, multiple TTS platforms support high-quality and human-alike speech generation~\cite{playht, elevenlabs}.

Although voice synthesis is rapidly developing, it remains susceptible to adversarial examples, as it is fundamentally based on deep learning technology. 
Attack-VC~\cite{huang2021defending} is the first work leveraging adversarial examples to spoof voice synthesis models, misleading them into producing unqualified voice samples. 
Despite its effectiveness on the target voice synthesis model, this approach requires long computation time to optimize adversarial perturbations, making it impractical for real-time speech protection.
VSMask~\cite{wang2023vsmask} attempts to address this issue by designing a predictive model that forecasts perturbations for upcoming streaming speech. 
While it overcomes the latency issue caused by offline optimization, the approach still depends on gradient back-propagation to optimize the predictive model, degrading its black-box performance.
SampleMask~\cite{liu2023protecting} employs different types of frequency masks to protect speech samples while preserving their quality and achieves black-box protection by querying voice synthesis models.
However, its protection is limited in transferability, as it targets only a single model. 
Furthermore, the protection is not foolproof, leaving some samples vulnerable to attacks if the attacker attempts multiple times.
AntiFake~\cite{yu2023antifake} offers effective and transferable protection across various voice synthesis models. 
Nevertheless, since it injects optimized noise into the raw speech sample, it struggles to protect samples in real-time while maintaining their naturalness and clarity.
In contrast, \ours overcomes the limitations of existing methods with a noise-free and universal protection mechanism, providing a robust solution for both online and offline applications.
\vspace{-10pt}
\subsection{Audio Style Transfer}
Audio styles, characterized by elements such as timbre, spatialization, and loudness, present complex variables that are challenging to manually manipulate.
Recent developments in automatic audio generation models enable intelligent audio style transfer with high performance and efficiency. 
Drawing inspiration from image style transfer~\cite{gatys2016image}, Grinstein et al.~\cite{grinstein2018audio} propose an innovative audio style transfer framework utilizing convolutional neural networks.
In addition, the approach of differentiable signal processing marks a significant enhancement in this field~\cite{ramirez2021differentiable}, optimizing the audio style transfer framework through backpropagation and effectively addressing the issue of weak correlation between audio and parameters.
DeepAFx-ST~\cite{steinmetz2022style} applies differentiable signal processing along with self-supervised training, achieving audio style transfer without reliance on labeled data.
Meanwhile, SpeechSplit~\cite{qian2020unsupervised} provides a speech decomposition method to separate speech into four components: content, pitch, rhythm, and timbre.
This method enables recent TTS models to generate speech with various audio styles~\cite{huang2022generspeech}.

Moreover, style transfer can be utilized for adversarial example generation.
For example, StyleFool~\cite{cao2023stylefool} presents an attack based on style transfer to fool video classification models.
Jin et al.~\cite{jin2024towards} utilize style-transferred speech audios to attack speech recognition models.
SMACK~\cite{yu2023smack} successfully compromise speech recognition models by modifying the prosodies of benign speech samples.
In this work, we apply audio style transfer to protect human speech against voice deepfake attacks without introducing additional noise.

\vspace{-8pt}
\subsection{Reverberation Applications} 
Reverberation is caused by sound waves reflecting off surfaces in an environment.
Reverberation effects are particularly noticeable in large indoor environments, where the sound persists after the original sound stops.
Haas~\cite{haas1972influence} conducts a comprehensive study on the impact of reverberation on human hearing, noting that humans cannot distinguish reverberations that diminish quickly.
Although reverberation is usually very weak, it is not negligible in audio signal processing.
A data augmentation approach based on reverberation simulation is used for speech~\cite{ko2017study} or speaker recognition models~\cite{lin2022robust}, improving recognition accuracy, especially in far-field scenarios.
Meanwhile, reverberation plays a crucial role in over-the-air adversarial audio attacks. 
Since adversarial speech samples are created by adding subtle perturbations to the original audio, which are weak and easily affected by reverberation, these attacks often fail to compromise speech recognition models in the physical world.
To overcome this weakness, Imperio~\cite{schonherr2020imperio} designs a solution to estimate the RIR in the room and adjust the initial adversarial audio signal, while Chen et al.~\cite{chen2020metamorph} further improve the effective attack range through channel state estimation.
Recent work points out that reverberation can also be used for adversarial audio signal generation.
For example, AdvReverb~\cite{chen2023advreverb} designs an adversarial audio attack by adding reverberation, and TrojanRoom~\cite{chendevil24} proposes a new attack using RIR to trigger a backdoor in speech recognition without injecting extra noise.
In \ours, we employ sophisticated manipulation of natural RIR to generate reverberation in speech, providing protection against voice deepfake attacks while preserving speech naturalness and quality.

\section{Conclusion}
The misuse of voice synthesis technology presents a significant threat to voice data security and privacy. 
Although numerous defense mechanisms have been proposed in previous studies, they often demonstrate limited effectiveness in real-world applications. 
In this paper, we introduce \ours, a noise-free defense method designed to mitigate voice deepfake attacks by unknown adversaries. 
Additionally, we propose \ourss, a real-time protection mode of \ours, which is intended for instant communication applications. 
We evaluate \ours and \ourss against various unseen open-source and commercial voice cloning models. 
The experimental results demonstrate that \ours and \ourss can effectively prevent synthesized voice from deceiving ASV models or human ears while preserving speech clarity and naturalness. 
In summary, \ours is the first defense mechanism to successfully integrate effectiveness, transferability, naturalness, and real-time capability in combating deepfake voice generation.

\begin{acks}
We would like to extend our appreciation to the shepherd and
anonymous reviewers for their invaluable input on our study.
This work was supported in part by the U.S. National Science
Foundation grant CNS-2310207.
\end{acks}

\bibliographystyle{ACM-Reference-Format}
\bibliography{reference}


\begin{thebibliography}{53}


\ifx \showCODEN    \undefined \def \showCODEN     #1{\unskip}     \fi
\ifx \showDOI      \undefined \def \showDOI       #1{#1}\fi
\ifx \showISBNx    \undefined \def \showISBNx     #1{\unskip}     \fi
\ifx \showISBNxiii \undefined \def \showISBNxiii  #1{\unskip}     \fi
\ifx \showISSN     \undefined \def \showISSN      #1{\unskip}     \fi
\ifx \showLCCN     \undefined \def \showLCCN      #1{\unskip}     \fi
\ifx \shownote     \undefined \def \shownote      #1{#1}          \fi
\ifx \showarticletitle \undefined \def \showarticletitle #1{#1}   \fi
\ifx \showURL      \undefined \def \showURL       {\relax}        \fi
\providecommand\bibfield[2]{#2}
\providecommand\bibinfo[2]{#2}
\providecommand\natexlab[1]{#1}
\providecommand\showeprint[2][]{arXiv:#2}

\bibitem[fac(2010)]%
        {facetime}
 \bibinfo{year}{2010}\natexlab{}.
\newblock \bibinfo{title}{Apple FaceTime}.
\newblock \bibinfo{howpublished}{\url{https://apps.apple.com/us/app/facetime/}}.
\newblock


\bibitem[sca(2019)]%
        {scammoney}
 \bibinfo{year}{2019}\natexlab{}.
\newblock \bibinfo{title}{Fraudsters Used AI to Mimic CEO’s Voice in Unusual Cybercrime Case}.
\newblock \bibinfo{howpublished}{\url{https://www.wsj.com/articles/fraudsters-use-ai-to-mimic-ceos-voice-in-unusual-cybercrime-case-11567157402}}.
\newblock


\bibitem[ele(2023)]%
        {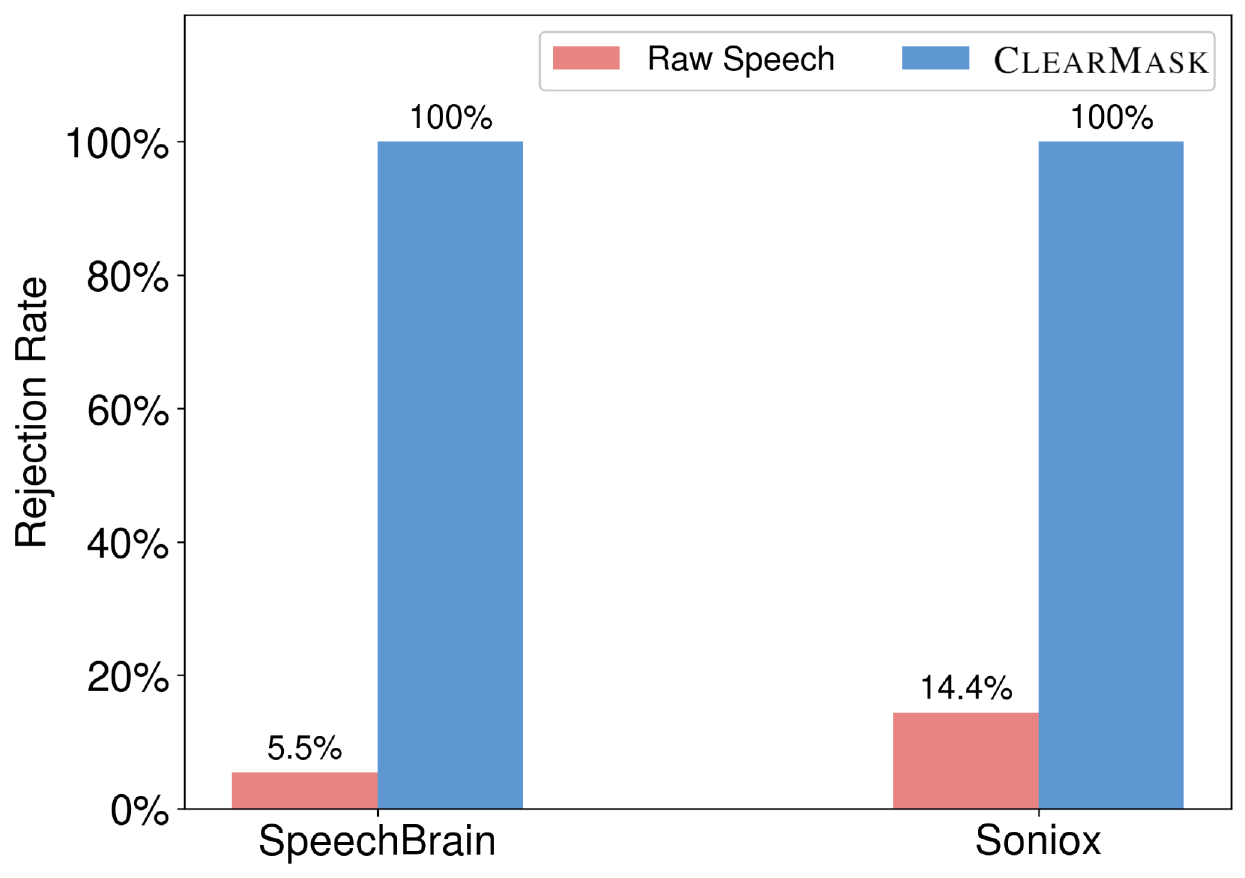}
 \bibinfo{year}{2023}\natexlab{}.
\newblock \bibinfo{title}{ElevenLabs}.
\newblock \bibinfo{howpublished}{\url{https://www.elevenlabs.io/}}.
\newblock


\bibitem[bre(2023)]%
        {breakbank}
 \bibinfo{year}{2023}\natexlab{}.
\newblock \bibinfo{title}{How I Broke Into a Bank Account With an AI-Generated Voice}.
\newblock \bibinfo{howpublished}{\url{https://www.vice.com/en/article/dy7axa/how-i-broke-into-a-bank-account-with-an-ai-generated-voice}}.
\newblock


\bibitem[pla(2023)]%
        {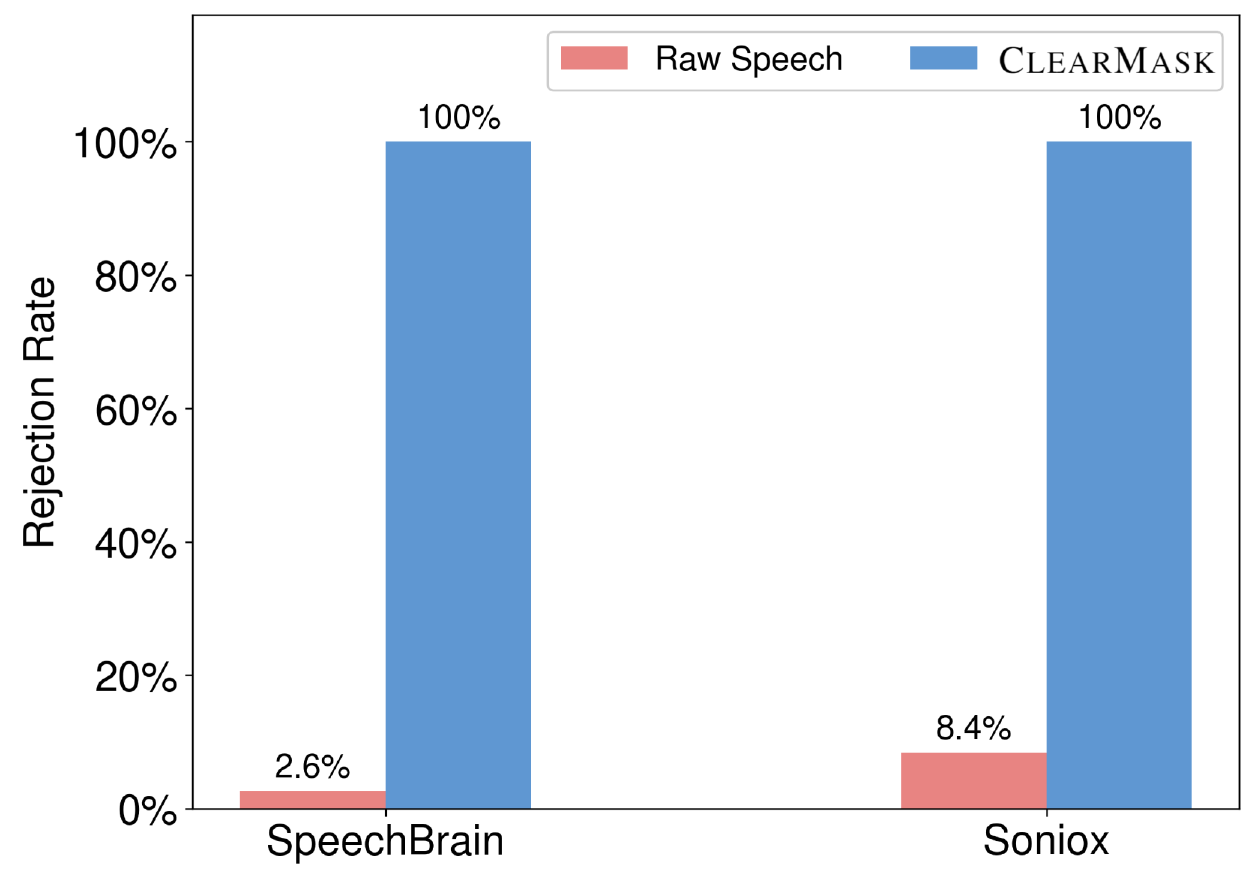}
 \bibinfo{year}{2023}\natexlab{}.
\newblock \bibinfo{title}{plat.ht}.
\newblock \bibinfo{howpublished}{\url{https://play.ht/}}.
\newblock


\bibitem[kid(2023)]%
        {kidnapping}
 \bibinfo{year}{2023}\natexlab{}.
\newblock \bibinfo{title}{‘Mom, these bad men have me’: She believes scammers cloned her daughter’s voice in a fake kidnapping}.
\newblock \bibinfo{howpublished}{\url{https://www.cnn.com/2023/04/29/us/ai-scam-calls-kidnapping-cec/index.html}}.
\newblock


\bibitem[Cha(2024)]%
        {ChatGPT}
 \bibinfo{year}{2024}\natexlab{}.
\newblock \bibinfo{title}{ChatGPT 4.0}.
\newblock \bibinfo{howpublished}{\url{https://chatgpt.com/}}.
\newblock


\bibitem[Goo(2024)]%
        {GoogleTTS}
 \bibinfo{year}{2024}\natexlab{}.
\newblock \bibinfo{title}{Google Text-to-Speech}.
\newblock \bibinfo{howpublished}{\url{https://cloud.google.com/text-to-speech}}.
\newblock


\bibitem[son(2024)]%
        {soniox}
 \bibinfo{year}{2024}\natexlab{}.
\newblock \bibinfo{title}{Soniox: Introducing AudioMind}.
\newblock \bibinfo{howpublished}{\url{https://soniox.com/}}.
\newblock


\bibitem[Ahmed et~al\mbox{.}(2020)]%
        {ahmed2020void}
\bibfield{author}{\bibinfo{person}{Muhammad~Ejaz Ahmed}, \bibinfo{person}{Il-Youp Kwak}, \bibinfo{person}{Jun~Ho Huh}, \bibinfo{person}{Iljoo Kim}, \bibinfo{person}{Taekkyung Oh}, {and} \bibinfo{person}{Hyoungshick Kim}.} \bibinfo{year}{2020}\natexlab{}.
\newblock \showarticletitle{Void: A fast and light voice liveness detection system}. In \bibinfo{booktitle}{\emph{29th USENIX Security Symposium (USENIX Security 20)}}. \bibinfo{pages}{2685--2702}.
\newblock


\bibitem[Ardila et~al\mbox{.}(2019)]%
        {ardila2019common}
\bibfield{author}{\bibinfo{person}{Rosana Ardila}, \bibinfo{person}{Megan Branson}, \bibinfo{person}{Kelly Davis}, \bibinfo{person}{Michael Henretty}, \bibinfo{person}{Michael Kohler}, \bibinfo{person}{Josh Meyer}, \bibinfo{person}{Reuben Morais}, \bibinfo{person}{Lindsay Saunders}, \bibinfo{person}{Francis~M Tyers}, {and} \bibinfo{person}{Gregor Weber}.} \bibinfo{year}{2019}\natexlab{}.
\newblock \showarticletitle{Common voice: A massively-multilingual speech corpus}.
\newblock \bibinfo{journal}{\emph{arXiv preprint arXiv:1912.06670}} (\bibinfo{year}{2019}).
\newblock


\bibitem[Cao et~al\mbox{.}(2023)]%
        {cao2023stylefool}
\bibfield{author}{\bibinfo{person}{Yuxin Cao}, \bibinfo{person}{Xi Xiao}, \bibinfo{person}{Ruoxi Sun}, \bibinfo{person}{Derui Wang}, \bibinfo{person}{Minhui Xue}, {and} \bibinfo{person}{Sheng Wen}.} \bibinfo{year}{2023}\natexlab{}.
\newblock \showarticletitle{Stylefool: Fooling video classification systems via style transfer}. In \bibinfo{booktitle}{\emph{2023 IEEE symposium on security and privacy (SP)}}. IEEE, \bibinfo{pages}{1631--1648}.
\newblock


\bibitem[Casanova et~al\mbox{.}(2022)]%
        {casanova2022yourtts}
\bibfield{author}{\bibinfo{person}{Edresson Casanova}, \bibinfo{person}{Julian Weber}, \bibinfo{person}{Christopher~D Shulby}, \bibinfo{person}{Arnaldo~Candido Junior}, \bibinfo{person}{Eren G{\"o}lge}, {and} \bibinfo{person}{Moacir~A Ponti}.} \bibinfo{year}{2022}\natexlab{}.
\newblock \showarticletitle{Yourtts: Towards zero-shot multi-speaker tts and zero-shot voice conversion for everyone}. In \bibinfo{booktitle}{\emph{International Conference on Machine Learning}}. PMLR, \bibinfo{pages}{2709--2720}.
\newblock


\bibitem[Chen et~al\mbox{.}(2023)]%
        {chen2023advreverb}
\bibfield{author}{\bibinfo{person}{Meng Chen}, \bibinfo{person}{Li Lu}, \bibinfo{person}{Jiadi Yu}, \bibinfo{person}{Zhongjie Ba}, \bibinfo{person}{Feng Lin}, {and} \bibinfo{person}{Kui Ren}.} \bibinfo{year}{2023}\natexlab{}.
\newblock \showarticletitle{AdvReverb: Rethinking the Stealthiness of Audio Adversarial Examples to Human Perception}.
\newblock \bibinfo{journal}{\emph{IEEE Transactions on Information Forensics and Security}} (\bibinfo{year}{2023}).
\newblock


\bibitem[Chen et~al\mbox{.}(2024)]%
        {chendevil24}
\bibfield{author}{\bibinfo{person}{Meng Chen}, \bibinfo{person}{Xiangyu Xu}, \bibinfo{person}{Li Lu}, \bibinfo{person}{Zhongjie Ba}, \bibinfo{person}{Feng Lin}, {and} \bibinfo{person}{Kui Ren}.} \bibinfo{year}{2024}\natexlab{}.
\newblock \showarticletitle{Devil in the Room: Triggering Audio Backdoors in the Physical World}. In \bibinfo{booktitle}{\emph{33th USENIX security symposium (USENIX Security 24)}}.
\newblock


\bibitem[Chen et~al\mbox{.}(2020)]%
        {chen2020metamorph}
\bibfield{author}{\bibinfo{person}{Tao Chen}, \bibinfo{person}{Longfei Shangguan}, \bibinfo{person}{Zhenjiang Li}, {and} \bibinfo{person}{Kyle Jamieson}.} \bibinfo{year}{2020}\natexlab{}.
\newblock \showarticletitle{Metamorph: Injecting inaudible commands into over-the-air voice controlled systems}. In \bibinfo{booktitle}{\emph{Network and Distributed Systems Security (NDSS) Symposium}}.
\newblock


\bibitem[Chen et~al\mbox{.}(2003)]%
        {chen2003voice}
\bibfield{author}{\bibinfo{person}{Yining Chen}, \bibinfo{person}{Min Chu}, \bibinfo{person}{Eric Chang}, \bibinfo{person}{Jia Liu}, {and} \bibinfo{person}{Runsheng Liu}.} \bibinfo{year}{2003}\natexlab{}.
\newblock \showarticletitle{Voice conversion with smoothed GMM and MAP adaptation}. In \bibinfo{booktitle}{\emph{Eighth European Conference on Speech Communication and Technology}}.
\newblock


\bibitem[Chen et~al\mbox{.}(2021)]%
        {chen2021again}
\bibfield{author}{\bibinfo{person}{Yen-Hao Chen}, \bibinfo{person}{Da-Yi Wu}, \bibinfo{person}{Tsung-Han Wu}, {and} \bibinfo{person}{Hung-yi Lee}.} \bibinfo{year}{2021}\natexlab{}.
\newblock \showarticletitle{Again-vc: A one-shot voice conversion using activation guidance and adaptive instance normalization}. In \bibinfo{booktitle}{\emph{ICASSP 2021-2021 IEEE International Conference on Acoustics, Speech and Signal Processing (ICASSP)}}. IEEE, \bibinfo{pages}{5954--5958}.
\newblock


\bibitem[Chou et~al\mbox{.}(2019)]%
        {chou2019one}
\bibfield{author}{\bibinfo{person}{Ju-chieh Chou}, \bibinfo{person}{Cheng-chieh Yeh}, {and} \bibinfo{person}{Hung-yi Lee}.} \bibinfo{year}{2019}\natexlab{}.
\newblock \showarticletitle{One-shot voice conversion by separating speaker and content representations with instance normalization}.
\newblock \bibinfo{journal}{\emph{arXiv preprint arXiv:1904.05742}} (\bibinfo{year}{2019}).
\newblock


\bibitem[Desplanques et~al\mbox{.}(2020)]%
        {desplanques2020ecapa}
\bibfield{author}{\bibinfo{person}{Brecht Desplanques}, \bibinfo{person}{Jenthe Thienpondt}, {and} \bibinfo{person}{Kris Demuynck}.} \bibinfo{year}{2020}\natexlab{}.
\newblock \showarticletitle{Ecapa-tdnn: Emphasized channel attention, propagation and aggregation in tdnn based speaker verification}.
\newblock \bibinfo{journal}{\emph{arXiv preprint arXiv:2005.07143}} (\bibinfo{year}{2020}).
\newblock


\bibitem[Gatys et~al\mbox{.}(2016)]%
        {gatys2016image}
\bibfield{author}{\bibinfo{person}{Leon~A Gatys}, \bibinfo{person}{Alexander~S Ecker}, {and} \bibinfo{person}{Matthias Bethge}.} \bibinfo{year}{2016}\natexlab{}.
\newblock \showarticletitle{Image style transfer using convolutional neural networks}. In \bibinfo{booktitle}{\emph{Proceedings of the IEEE conference on computer vision and pattern recognition}}. \bibinfo{pages}{2414--2423}.
\newblock


\bibitem[Grinstein et~al\mbox{.}(2018)]%
        {grinstein2018audio}
\bibfield{author}{\bibinfo{person}{Eric Grinstein}, \bibinfo{person}{Ngoc~QK Duong}, \bibinfo{person}{Alexey Ozerov}, {and} \bibinfo{person}{Patrick P{\'e}rez}.} \bibinfo{year}{2018}\natexlab{}.
\newblock \showarticletitle{Audio style transfer}. In \bibinfo{booktitle}{\emph{2018 IEEE international conference on acoustics, speech and signal processing (ICASSP)}}. IEEE, \bibinfo{pages}{586--590}.
\newblock


\bibitem[Haas(1972)]%
        {haas1972influence}
\bibfield{author}{\bibinfo{person}{Helmut Haas}.} \bibinfo{year}{1972}\natexlab{}.
\newblock \showarticletitle{The influence of a single echo on the audibility of speech}.
\newblock \bibinfo{journal}{\emph{Journal of the audio engineering society}} \bibinfo{volume}{20}, \bibinfo{number}{2} (\bibinfo{year}{1972}), \bibinfo{pages}{146--159}.
\newblock


\bibitem[Huang et~al\mbox{.}(2021)]%
        {huang2021defending}
\bibfield{author}{\bibinfo{person}{Chien-yu Huang}, \bibinfo{person}{Yist~Y Lin}, \bibinfo{person}{Hung-yi Lee}, {and} \bibinfo{person}{Lin-shan Lee}.} \bibinfo{year}{2021}\natexlab{}.
\newblock \showarticletitle{Defending your voice: Adversarial attack on voice conversion}. In \bibinfo{booktitle}{\emph{2021 IEEE Spoken Language Technology Workshop (SLT)}}. IEEE, \bibinfo{pages}{552--559}.
\newblock


\bibitem[Huang et~al\mbox{.}(2022)]%
        {huang2022generspeech}
\bibfield{author}{\bibinfo{person}{Rongjie Huang}, \bibinfo{person}{Yi Ren}, \bibinfo{person}{Jinglin Liu}, \bibinfo{person}{Chenye Cui}, {and} \bibinfo{person}{Zhou Zhao}.} \bibinfo{year}{2022}\natexlab{}.
\newblock \showarticletitle{Generspeech: Towards style transfer for generalizable out-of-domain text-to-speech}.
\newblock \bibinfo{journal}{\emph{Advances in Neural Information Processing Systems}}  \bibinfo{volume}{35} (\bibinfo{year}{2022}), \bibinfo{pages}{10970--10983}.
\newblock


\bibitem[Hussain et~al\mbox{.}(2021)]%
        {hussain2021waveguard}
\bibfield{author}{\bibinfo{person}{Shehzeen Hussain}, \bibinfo{person}{Paarth Neekhara}, \bibinfo{person}{Shlomo Dubnov}, \bibinfo{person}{Julian McAuley}, {and} \bibinfo{person}{Farinaz Koushanfar}.} \bibinfo{year}{2021}\natexlab{}.
\newblock \showarticletitle{$\{$WaveGuard$\}$: Understanding and mitigating audio adversarial examples}. In \bibinfo{booktitle}{\emph{30th USENIX security symposium (USENIX Security 21)}}. \bibinfo{pages}{2273--2290}.
\newblock


\bibitem[Jia et~al\mbox{.}(2018)]%
        {jia2018transfer}
\bibfield{author}{\bibinfo{person}{Ye Jia}, \bibinfo{person}{Yu Zhang}, \bibinfo{person}{Ron Weiss}, \bibinfo{person}{Quan Wang}, \bibinfo{person}{Jonathan Shen}, \bibinfo{person}{Fei Ren}, \bibinfo{person}{Patrick Nguyen}, \bibinfo{person}{Ruoming Pang}, \bibinfo{person}{Ignacio Lopez~Moreno}, \bibinfo{person}{Yonghui Wu}, {et~al\mbox{.}}} \bibinfo{year}{2018}\natexlab{}.
\newblock \showarticletitle{Transfer learning from speaker verification to multispeaker text-to-speech synthesis}.
\newblock \bibinfo{journal}{\emph{Advances in neural information processing systems}}  \bibinfo{volume}{31} (\bibinfo{year}{2018}).
\newblock


\bibitem[Jin et~al\mbox{.}(2024)]%
        {jin2024towards}
\bibfield{author}{\bibinfo{person}{Weifei Jin}, \bibinfo{person}{Yuxin Cao}, \bibinfo{person}{Junjie Su}, \bibinfo{person}{Qi Shen}, \bibinfo{person}{Kai Ye}, \bibinfo{person}{Derui Wang}, \bibinfo{person}{Jie Hao}, {and} \bibinfo{person}{Ziyao Liu}.} \bibinfo{year}{2024}\natexlab{}.
\newblock \showarticletitle{Towards Evaluating the Robustness of Automatic Speech Recognition Systems via Audio Style Transfer}. In \bibinfo{booktitle}{\emph{Proceedings of the 2nd ACM Workshop on Secure and Trustworthy Deep Learning Systems}}. \bibinfo{pages}{47--55}.
\newblock


\bibitem[Ko et~al\mbox{.}(2017)]%
        {ko2017study}
\bibfield{author}{\bibinfo{person}{Tom Ko}, \bibinfo{person}{Vijayaditya Peddinti}, \bibinfo{person}{Daniel Povey}, \bibinfo{person}{Michael~L Seltzer}, {and} \bibinfo{person}{Sanjeev Khudanpur}.} \bibinfo{year}{2017}\natexlab{}.
\newblock \showarticletitle{A study on data augmentation of reverberant speech for robust speech recognition}. In \bibinfo{booktitle}{\emph{2017 IEEE international conference on acoustics, speech and signal processing (ICASSP)}}. IEEE, \bibinfo{pages}{5220--5224}.
\newblock


\bibitem[Kong et~al\mbox{.}(2020)]%
        {kong2020hifi}
\bibfield{author}{\bibinfo{person}{Jungil Kong}, \bibinfo{person}{Jaehyeon Kim}, {and} \bibinfo{person}{Jaekyoung Bae}.} \bibinfo{year}{2020}\natexlab{}.
\newblock \showarticletitle{Hifi-gan: Generative adversarial networks for efficient and high fidelity speech synthesis}.
\newblock \bibinfo{journal}{\emph{Advances in neural information processing systems}}  \bibinfo{volume}{33} (\bibinfo{year}{2020}), \bibinfo{pages}{17022--17033}.
\newblock


\bibitem[Lin and Mak(2022)]%
        {lin2022robust}
\bibfield{author}{\bibinfo{person}{Weiwei Lin} {and} \bibinfo{person}{Man-Wai Mak}.} \bibinfo{year}{2022}\natexlab{}.
\newblock \showarticletitle{Robust speaker verification using population-based data augmentation}. In \bibinfo{booktitle}{\emph{ICASSP 2022-2022 IEEE International Conference on Acoustics, Speech and Signal Processing (ICASSP)}}. IEEE, \bibinfo{pages}{7642--7646}.
\newblock


\bibitem[Liu et~al\mbox{.}(2023)]%
        {liu2023protecting}
\bibfield{author}{\bibinfo{person}{Zihao Liu}, \bibinfo{person}{Yan Zhang}, {and} \bibinfo{person}{Chenglin Miao}.} \bibinfo{year}{2023}\natexlab{}.
\newblock \showarticletitle{Protecting Your Voice from Speech Synthesis Attacks}. In \bibinfo{booktitle}{\emph{Proceedings of the 39th Annual Computer Security Applications Conference}}. \bibinfo{pages}{394--408}.
\newblock


\bibitem[Madry et~al\mbox{.}(2018)]%
        {madry2018towards}
\bibfield{author}{\bibinfo{person}{Aleksander Madry}, \bibinfo{person}{Aleksandar Makelov}, \bibinfo{person}{Ludwig Schmidt}, \bibinfo{person}{Dimitris Tsipras}, {and} \bibinfo{person}{Adrian Vladu}.} \bibinfo{year}{2018}\natexlab{}.
\newblock \showarticletitle{Towards Deep Learning Models Resistant to Adversarial Attacks}. In \bibinfo{booktitle}{\emph{International Conference on Learning Representations}}.
\newblock


\bibitem[Panayotov et~al\mbox{.}(2015)]%
        {7178964}
\bibfield{author}{\bibinfo{person}{Vassil Panayotov}, \bibinfo{person}{Guoguo Chen}, \bibinfo{person}{Daniel Povey}, {and} \bibinfo{person}{Sanjeev Khudanpur}.} \bibinfo{year}{2015}\natexlab{}.
\newblock \showarticletitle{Librispeech: An ASR corpus based on public domain audio books}. In \bibinfo{booktitle}{\emph{2015 IEEE International Conference on Acoustics, Speech and Signal Processing (ICASSP)}}. \bibinfo{pages}{5206--5210}.
\newblock
\urldef\tempurl%
\url{https://doi.org/10.1109/ICASSP.2015.7178964}
\showDOI{\tempurl}


\bibitem[Paszke et~al\mbox{.}(2019)]%
        {paszke2019pytorch}
\bibfield{author}{\bibinfo{person}{Adam Paszke}, \bibinfo{person}{Sam Gross}, \bibinfo{person}{Francisco Massa}, \bibinfo{person}{Adam Lerer}, \bibinfo{person}{James Bradbury}, \bibinfo{person}{Gregory Chanan}, \bibinfo{person}{Trevor Killeen}, \bibinfo{person}{Zeming Lin}, \bibinfo{person}{Natalia Gimelshein}, \bibinfo{person}{Luca Antiga}, {et~al\mbox{.}}} \bibinfo{year}{2019}\natexlab{}.
\newblock \showarticletitle{Pytorch: An imperative style, high-performance deep learning library}.
\newblock \bibinfo{journal}{\emph{Advances in neural information processing systems}}  \bibinfo{volume}{32} (\bibinfo{year}{2019}).
\newblock


\bibitem[Popov et~al\mbox{.}(2021)]%
        {popov2021diffusion}
\bibfield{author}{\bibinfo{person}{Vadim Popov}, \bibinfo{person}{Ivan Vovk}, \bibinfo{person}{Vladimir Gogoryan}, \bibinfo{person}{Tasnima Sadekova}, \bibinfo{person}{Mikhail Kudinov}, {and} \bibinfo{person}{Jiansheng Wei}.} \bibinfo{year}{2021}\natexlab{}.
\newblock \showarticletitle{Diffusion-based voice conversion with fast maximum likelihood sampling scheme}.
\newblock \bibinfo{journal}{\emph{arXiv preprint arXiv:2109.13821}} (\bibinfo{year}{2021}).
\newblock


\bibitem[Qian et~al\mbox{.}(2020)]%
        {qian2020unsupervised}
\bibfield{author}{\bibinfo{person}{Kaizhi Qian}, \bibinfo{person}{Yang Zhang}, \bibinfo{person}{Shiyu Chang}, \bibinfo{person}{Mark Hasegawa-Johnson}, {and} \bibinfo{person}{David Cox}.} \bibinfo{year}{2020}\natexlab{}.
\newblock \showarticletitle{Unsupervised speech decomposition via triple information bottleneck}. In \bibinfo{booktitle}{\emph{International Conference on Machine Learning}}. PMLR, \bibinfo{pages}{7836--7846}.
\newblock


\bibitem[Qian et~al\mbox{.}(2019)]%
        {qian2019autovc}
\bibfield{author}{\bibinfo{person}{Kaizhi Qian}, \bibinfo{person}{Yang Zhang}, \bibinfo{person}{Shiyu Chang}, \bibinfo{person}{Xuesong Yang}, {and} \bibinfo{person}{Mark Hasegawa-Johnson}.} \bibinfo{year}{2019}\natexlab{}.
\newblock \showarticletitle{Autovc: Zero-shot voice style transfer with only autoencoder loss}. In \bibinfo{booktitle}{\emph{International Conference on Machine Learning}}. PMLR, \bibinfo{pages}{5210--5219}.
\newblock


\bibitem[Ram{\'\i}rez et~al\mbox{.}(2021)]%
        {ramirez2021differentiable}
\bibfield{author}{\bibinfo{person}{Marco A~Mart{\'\i}nez Ram{\'\i}rez}, \bibinfo{person}{Oliver Wang}, \bibinfo{person}{Paris Smaragdis}, {and} \bibinfo{person}{Nicholas~J Bryan}.} \bibinfo{year}{2021}\natexlab{}.
\newblock \showarticletitle{Differentiable signal processing with black-box audio effects}. In \bibinfo{booktitle}{\emph{ICASSP 2021-2021 IEEE International Conference on Acoustics, Speech and Signal Processing (ICASSP)}}. IEEE, \bibinfo{pages}{66--70}.
\newblock


\bibitem[Ravanelli et~al\mbox{.}(2021)]%
        {speechbrain}
\bibfield{author}{\bibinfo{person}{Mirco Ravanelli}, \bibinfo{person}{Titouan Parcollet}, \bibinfo{person}{Peter Plantinga}, \bibinfo{person}{Aku Rouhe}, \bibinfo{person}{Samuele Cornell}, \bibinfo{person}{Loren Lugosch}, \bibinfo{person}{Cem Subakan}, \bibinfo{person}{Nauman Dawalatabad}, \bibinfo{person}{Abdelwahab Heba}, \bibinfo{person}{Jianyuan Zhong}, \bibinfo{person}{Ju-Chieh Chou}, \bibinfo{person}{Sung-Lin Yeh}, \bibinfo{person}{Szu-Wei Fu}, \bibinfo{person}{Chien-Feng Liao}, \bibinfo{person}{Elena Rastorgueva}, \bibinfo{person}{François Grondin}, \bibinfo{person}{William Aris}, \bibinfo{person}{Hwidong Na}, \bibinfo{person}{Yan Gao}, \bibinfo{person}{Renato~De Mori}, {and} \bibinfo{person}{Yoshua Bengio}.} \bibinfo{year}{2021}\natexlab{}.
\newblock \bibinfo{title}{{SpeechBrain}: A General-Purpose Speech Toolkit}.
\newblock
\newblock
\showeprint[arxiv]{2106.04624}~[eess.AS]
\newblock
\shownote{arXiv:2106.04624}.


\bibitem[Sch{\"o}nherr et~al\mbox{.}(2020)]%
        {schonherr2020imperio}
\bibfield{author}{\bibinfo{person}{Lea Sch{\"o}nherr}, \bibinfo{person}{Thorsten Eisenhofer}, \bibinfo{person}{Steffen Zeiler}, \bibinfo{person}{Thorsten Holz}, {and} \bibinfo{person}{Dorothea Kolossa}.} \bibinfo{year}{2020}\natexlab{}.
\newblock \showarticletitle{Imperio: Robust over-the-air adversarial examples for automatic speech recognition systems}. In \bibinfo{booktitle}{\emph{Proceedings of the 36th Annual Computer Security Applications Conference}}. \bibinfo{pages}{843--855}.
\newblock


\bibitem[Steinmetz et~al\mbox{.}(2022)]%
        {steinmetz2022style}
\bibfield{author}{\bibinfo{person}{Christian~J Steinmetz}, \bibinfo{person}{Nicholas~J Bryan}, {and} \bibinfo{person}{Joshua~D Reiss}.} \bibinfo{year}{2022}\natexlab{}.
\newblock \showarticletitle{Style transfer of audio effects with differentiable signal processing}.
\newblock \bibinfo{journal}{\emph{arXiv preprint arXiv:2207.08759}} (\bibinfo{year}{2022}).
\newblock


\bibitem[Traer and McDermott(2016)]%
        {traer2016statistics}
\bibfield{author}{\bibinfo{person}{James Traer} {and} \bibinfo{person}{Josh~H McDermott}.} \bibinfo{year}{2016}\natexlab{}.
\newblock \showarticletitle{Statistics of natural reverberation enable perceptual separation of sound and space}.
\newblock \bibinfo{journal}{\emph{Proceedings of the National Academy of Sciences}} \bibinfo{volume}{113}, \bibinfo{number}{48} (\bibinfo{year}{2016}), \bibinfo{pages}{E7856--E7865}.
\newblock


\bibitem[Veaux et~al\mbox{.}(2016)]%
        {veaux2016superseded}
\bibfield{author}{\bibinfo{person}{Christophe Veaux}, \bibinfo{person}{Junichi Yamagishi}, \bibinfo{person}{Kirsten MacDonald}, {et~al\mbox{.}}} \bibinfo{year}{2016}\natexlab{}.
\newblock \showarticletitle{Superseded-cstr vctk corpus: English multi-speaker corpus for cstr voice cloning toolkit}.
\newblock  (\bibinfo{year}{2016}).
\newblock


\bibitem[Wan et~al\mbox{.}(2018)]%
        {wan2018generalized}
\bibfield{author}{\bibinfo{person}{Li Wan}, \bibinfo{person}{Quan Wang}, \bibinfo{person}{Alan Papir}, {and} \bibinfo{person}{Ignacio~Lopez Moreno}.} \bibinfo{year}{2018}\natexlab{}.
\newblock \showarticletitle{Generalized end-to-end loss for speaker verification}. In \bibinfo{booktitle}{\emph{2018 IEEE International Conference on Acoustics, Speech and Signal Processing (ICASSP)}}. IEEE, \bibinfo{pages}{4879--4883}.
\newblock


\bibitem[Wang et~al\mbox{.}(2023a)]%
        {wang2023vsmask}
\bibfield{author}{\bibinfo{person}{Yuanda Wang}, \bibinfo{person}{Hanqing Guo}, \bibinfo{person}{Guangjing Wang}, \bibinfo{person}{Bocheng Chen}, {and} \bibinfo{person}{Qiben Yan}.} \bibinfo{year}{2023}\natexlab{a}.
\newblock \showarticletitle{Vsmask: Defending against voice synthesis attack via real-time predictive perturbation}. In \bibinfo{booktitle}{\emph{Proceedings of the 16th ACM Conference on Security and Privacy in Wireless and Mobile Networks}}. \bibinfo{pages}{239--250}.
\newblock


\bibitem[Wang et~al\mbox{.}(2022)]%
        {wang2022ghosttalk}
\bibfield{author}{\bibinfo{person}{Yuanda Wang}, \bibinfo{person}{Hanqing Guo}, {and} \bibinfo{person}{Qiben Yan}.} \bibinfo{year}{2022}\natexlab{}.
\newblock \showarticletitle{GhostTalk: Interactive Attack on Smartphone Voice System Through Power Line}. In \bibinfo{booktitle}{\emph{Network and Distributed Systems Security (NDSS) Symposium}}.
\newblock


\bibitem[Wang et~al\mbox{.}(2018)]%
        {wang2018style}
\bibfield{author}{\bibinfo{person}{Yuxuan Wang}, \bibinfo{person}{Daisy Stanton}, \bibinfo{person}{Yu Zhang}, \bibinfo{person}{RJ-Skerry Ryan}, \bibinfo{person}{Eric Battenberg}, \bibinfo{person}{Joel Shor}, \bibinfo{person}{Ying Xiao}, \bibinfo{person}{Ye Jia}, \bibinfo{person}{Fei Ren}, {and} \bibinfo{person}{Rif~A Saurous}.} \bibinfo{year}{2018}\natexlab{}.
\newblock \showarticletitle{Style tokens: Unsupervised style modeling, control and transfer in end-to-end speech synthesis}. In \bibinfo{booktitle}{\emph{International conference on machine learning}}. PMLR, \bibinfo{pages}{5180--5189}.
\newblock


\bibitem[Wang et~al\mbox{.}(2023b)]%
        {wang2023practical}
\bibfield{author}{\bibinfo{person}{Yuanda Wang}, \bibinfo{person}{Qiben Yan}, \bibinfo{person}{Nikolay Ivanov}, {and} \bibinfo{person}{Xun Chen}.} \bibinfo{year}{2023}\natexlab{b}.
\newblock \showarticletitle{A Practical Survey on Emerging Threats from AI-driven Voice Attacks: How Vulnerable are Commercial Voice Control Systems?}
\newblock \bibinfo{journal}{\emph{arXiv preprint arXiv:2312.06010}} (\bibinfo{year}{2023}).
\newblock


\bibitem[Wenger et~al\mbox{.}(2021)]%
        {wenger2021hello}
\bibfield{author}{\bibinfo{person}{Emily Wenger}, \bibinfo{person}{Max Bronckers}, \bibinfo{person}{Christian Cianfarani}, \bibinfo{person}{Jenna Cryan}, \bibinfo{person}{Angela Sha}, \bibinfo{person}{Haitao Zheng}, {and} \bibinfo{person}{Ben~Y Zhao}.} \bibinfo{year}{2021}\natexlab{}.
\newblock \showarticletitle{" Hello, It's Me": Deep Learning-based Speech Synthesis Attacks in the Real World}. In \bibinfo{booktitle}{\emph{Proceedings of the 2021 ACM SIGSAC Conference on Computer and Communications Security}}. \bibinfo{pages}{235--251}.
\newblock


\bibitem[Yu et~al\mbox{.}(2023a)]%
        {yu2023smack}
\bibfield{author}{\bibinfo{person}{Zhiyuan Yu}, \bibinfo{person}{Yuanhaur Chang}, \bibinfo{person}{Ning Zhang}, {and} \bibinfo{person}{Chaowei Xiao}.} \bibinfo{year}{2023}\natexlab{a}.
\newblock \showarticletitle{$\{$SMACK$\}$: Semantically Meaningful Adversarial Audio Attack}. In \bibinfo{booktitle}{\emph{32nd USENIX Security Symposium (USENIX Security 23)}}. \bibinfo{pages}{3799--3816}.
\newblock


\bibitem[Yu et~al\mbox{.}(2023b)]%
        {yu2023antifake}
\bibfield{author}{\bibinfo{person}{Zhiyuan Yu}, \bibinfo{person}{Shixuan Zhai}, {and} \bibinfo{person}{Ning Zhang}.} \bibinfo{year}{2023}\natexlab{b}.
\newblock \showarticletitle{Antifake: Using adversarial audio to prevent unauthorized speech synthesis}. In \bibinfo{booktitle}{\emph{Proceedings of the 2023 ACM SIGSAC Conference on Computer and Communications Security}}. \bibinfo{pages}{460--474}.
\newblock


\bibitem[Zhadan(2023)]%
        {voiceabuse}
\bibfield{author}{\bibinfo{person}{Anna Zhadan}.} \bibinfo{year}{2023}\natexlab{}.
\newblock \bibinfo{title}{Emma Watson reads Mein Kampf while Biden announces invasion of Russia in latest AI voice clone abuse}.
\newblock \bibinfo{howpublished}{\url{https://cybernews.com/news/ai-voice-clone-misuse/}}.
\newblock


\end{thebibliography}

\appendix
\section*{Appendix}
\section{Voice Feature and Audio Style}\label{sec:voice_and_style}
\begin{figure}[htbp]
    \centering
    \subfigure[Voice Embedding Distribution]{\includegraphics[width=0.22\textwidth]{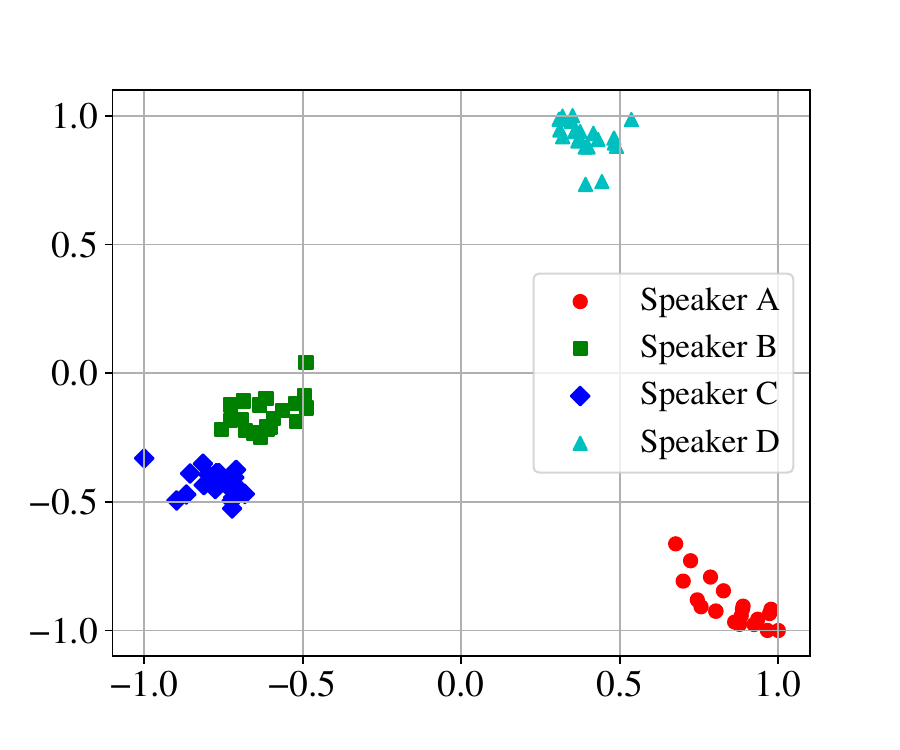}
    \label{fig: voice_distribution}}
    \
    \subfigure[Style Embedding Distribution]{\includegraphics[width=0.22\textwidth]{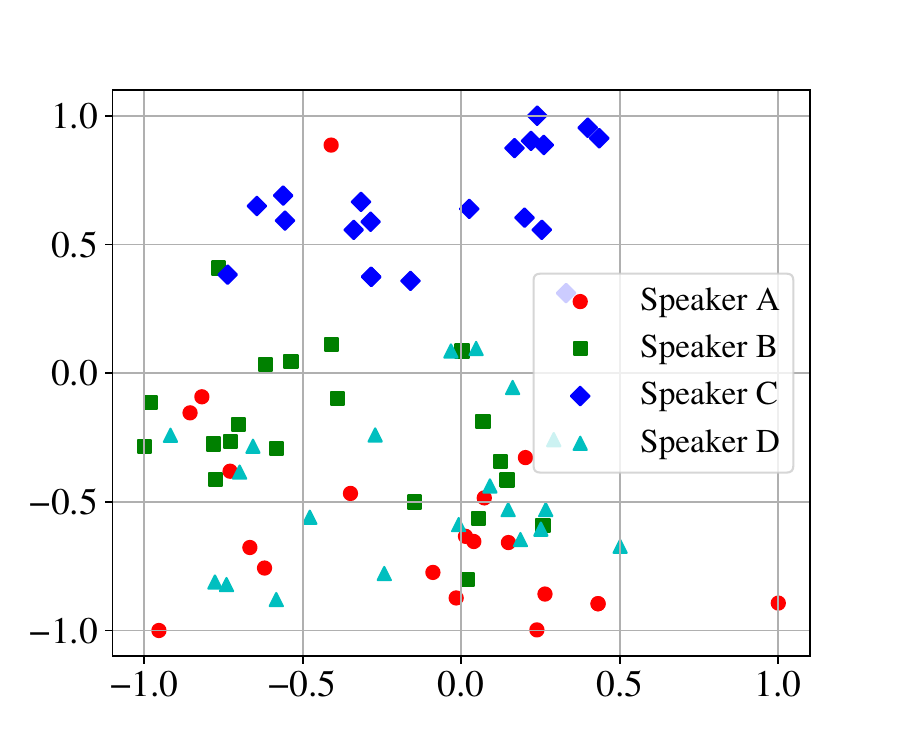}
    \label{fig: style_distribution}}
    \caption{Voice and audio style embedding vectors from the same speakers.}
    \label{fig: voice_and_style}
\end{figure}

When using audio style transfer to process speech audio, the most critical challenge is: \textit{how to select the target audio style for protection?}
A straightforward solution is to select an audio style from another speaker whose voice is different from the protected speaker, and apply it to the original speech.
However, this is impractical.
Fig.~\ref{fig: voice_and_style} shows the distributions of voice embedding vectors and audio style embedding vectors from the same speech samples but different speakers, with all vectors  are reduced in dimensionality using principal component analysis (PCA).
The voice embedding vectors exhibit clear boundaries between different speakers, while the audio style embeddings from different speakers are randomly mixed.
This is because voice encoders are only trained on human speech, whereas audio style features are derived from a wide range of sounds beyond human language, covering a broader spectrum of auditory characteristics.
As a result, audio style extraction cannot precisely distinguish between voice characteristics, suggesting that different voices may share similar audio styles.
This insight motivates us to explore new audio styles to generate effective adversarial examples.

\section{Textual Input Samples}\label{appendix_samples}
Table~\ref{Table: deepfake_voice_input} shows a few textual input samples for deepfake voice synthesis, which are generated via GPT-4.

\begin{table}[b]
\centering
\footnotesize
\caption{Textual input samples for deepfake voice synthesis.}
\label{Table: deepfake_voice_input}
\begin{tabular}{p{0.45\textwidth}} 
\toprule
\textbf{Sentences}\\ \midrule
Hi, I'm calling from HR. We're updating our records and need your Social Security number. \\ 
\midrule
This is the medical office. We require the patient's file to be sent to a new email immediately. \\ 
\midrule
Finance department here. We need the wire transfer code for the recent transaction to finalize it. \\ 
\midrule
This is tech support. We've detected unusual activity on your account. Please provide your password for verification. \\ 
\midrule
I'm outside the building without my ID. Can you buzz me in or give the access code? \\ 
\midrule
This is from IT support. We've noticed a security breach. Please provide your username and temporary password immediately. \\ 
\midrule
Hi, we're updating the security protocol. Can you confirm your employee ID and access badge number for verification? \\ 
\midrule
Hello, I'm handling a critical project update. Can you email me the latest financial forecast document right now? \\ 
\midrule
This is the service desk. To restore your account access, we need you to confirm your mother's maiden name and birth date. \\ 
\midrule
Calling from customer service. To prevent your account from being locked, please provide the recent one-time password sent to you. \\ 
\bottomrule
\end{tabular}
\end{table}

\end{document}